\documentclass[manuscript,screen,nonacm]{acmart}

\usepackage{booktabs}
\usepackage{array}
\usepackage{amsmath, amssymb}
\DeclareFontFamily{OMX}{cfex}{}
\DeclareFontShape{OMX}{cfex}{m}{n}{<->cmex10}{}
\DeclareSymbolFont{cfoperators}{OMX}{cfex}{m}{n}
\DeclareMathSymbol{\cfsum}{\mathop}{cfoperators}{"50}



\title{When Do LLMs Model the Right Target? Problem Formulation in
High-Stakes Risk Prediction}

\author{Jennifer Zhang}
\affiliation{%
  \department{Department of Statistics}
  \institution{University of California, Berkeley}
  \city{Berkeley}
  \state{California}
  \country{USA}
}
\author{Amanda Coston}
\affiliation{%
  \department{Department of Statistics}
  \institution{University of California, Berkeley}
  \city{Berkeley}
  \state{California}
  \country{USA}
}

\begin{document}

\begin{abstract}
Generative AI is increasingly used to assist with data science tasks, yet fundamental questions remain about its ability to correctly formulate analytical problems. We investigate how well large language models (LLMs) can formulate prediction problems when developing models for high-stakes decision making. In this setting, choices about what to predict and how to estimate it are critical to validity: whether a model estimates the intended target. We focus on one particular challenge to validity: decision dependency, which arises when historical decisions affect observed outcomes. Prior work has shown that failing to account for decision dependency can lead to invalid predictive models, but little is known about whether LLMs can appropriately address this challenge during model development. We evaluate this aspect of problem formulation in LLM-generated code and explanations across child-welfare, healthcare, and employment settings, using 11 models from OpenAI and Anthropic and seven prompt conditions. We assess whether the LLM-generated code implements a procedure appropriate for estimating the intended outcome under the study's assumptions and whether the explanation acknowledges decision dependency. We find that LLMs do not reliably address decision dependency without guidance in the prompt. When prompts specify the target, the assumption needed to estimate it, and the estimation procedure, target-aligned implementation reaches 99--100\%. We find that among frontier models, a relatively simple prompt description of how decisions affect outcomes appears sufficient to elicit target-aligned implementation. Some responses also acknowledge decision dependency without addressing it in their code. Our findings suggest that analysts should not rely on LLMs to correctly formulate prediction problems without explicit methodological guidance and careful review. We propose validity-centered guidelines for LLM-assisted predictive modeling, covering prompt formulation, validation of the target and estimation procedure, and identification of key assumptions. Establishing validity is a necessary early step in evaluating whether to deploy a predictive model in a high-stakes setting, but deployment requires weighing broader ethical, institutional, and practical considerations \cite{selbst2019fairness, metcalf2021algorithmic, green2022flaws, kawakami2024situate}.

\end{abstract}

\maketitle

\section{Introduction}
\label{sec:intro}

Generative AI is increasingly used to assist with data science tasks, from data visualization to hypothesis testing and predictive modeling \cite{o2025survey, dibia2023lida}.
In benchmark settings, LLMs have correctly generated code when the prompt specifies the problem formulation \cite{boominathan2026empirical}.  
Yet fundamental questions remain about whether LLMs can correctly formulate an analytical problem without prompt guidance \cite{o2025survey, gu2024analysts, song2026statllm, wang2026benchmarking}. These questions are particularly pressing given recent studies finding that analysts defer to LLM agents on problem formulation choices: in a recent study by Anthropic, 92\% of expert analysts selected the outcome measure their LLM agent was randomly instructed to favor \citep{herdagdelens,anthropic2026judgmentcalls}.

The question of proper problem formulation is particularly important for predictive modeling in high-stakes settings. Problem formulation choices---what outcome to predict, for whom, and using which data---impact \emph{validity}: whether the model meets its intended purpose \citep{passi2019problem, jacobs2021measurement}. A model can achieve high predictive accuracy yet remain invalid for its intended use \citep{guerdan2023groundless, coston2022validity, wang2023against}.
For instance, 
a widely deployed healthcare algorithm predicted healthcare expenditure as a proxy for health needs \citep{obermeyer2019dissecting}. 
The model appeared to be well-calibrated by race with respect to costs, but expenditures differentially reflect actual needs by race: at the same risk score, Black patients were considerably sicker than White patients \citep{obermeyer2019dissecting}. 
In child welfare, models trained to predict observed outcomes without accounting for \emph{decision dependencies}---that historical decisions impacted outcomes---mistakenly flag cases that were responsive to intervention as low-risk \cite{coston2020counterfactual}.
These examples show that a model can predict recorded outcomes accurately while estimating the wrong
quantity for the intended use.

Meanwhile, recent years have seen significant research interest in using LLMs for child welfare, clinical care, and other high-stakes decision-making settings \citep{moon2026promises, cruz2024evaluating, mallari2024generative, saveliev2024climb}. For instance, recent work proposed a system to let clinicians request predictive analyses in natural language, with an LLM agent guiding them through the full modeling pipeline \citep{saveliev2024climb, saveliev2025climbdc}.
In such workflows, the LLM is positioned to make problem formulation choices that determine whether the resulting model is fit for its purpose.
Existing work evaluating LLMs in high-stakes decision making has largely focused on predictive performance and fairness, rather than whether LLMs formulate prediction problems appropriately  \citep{lee2026auditing, mallari2024generative}.
This motivates our central question: can LLMs make problem formulation choices that align a predictive model with its intended use?

In this paper we investigate whether LLMs can address a particular problem formulation challenge, \emph{decision dependencies}, when generating predictive models for high-stakes decision making. In many high-stakes decision-making settings, the target of interest is the risk of an adverse outcome under a particular decision \cite{guerdan2023groundless,coston2022validity}. For example, the target may be a child's risk of harm or neglect if their case were screened out, or a patient's risk of poor health outcomes under standard care rather than high-risk care management. 
This target often differs from outcomes recorded in the data, which may have been observed under different decisions than the proposed decision. For instance, patients who receive more intensive care may have better observed outcomes because treatment reduced their future health risk, making them appear lower-risk in historical data \cite{caruana2015intelligible}.
Failure to account for these \emph{decision dependencies} risks building models that systematically screen out those responsive to increased intervention \cite{coston2020counterfactual}. We focus on decision dependencies because, unlike formulation challenges such as measurement error due to proxy outcomes, decision dependencies can be addressed under suitable conditions through a concrete design fix: restricting the training data to those who received the decision of interest \cite{coston2020counterfactual}. Decision dependencies thus serve as a concrete, testable case for evaluating LLMs' ability to appropriately formulate prediction problems.

\subsection{Study and contributions}

Analysts may provide different kinds of information when asking an LLM to build a predictive model. Studies of AI-assisted data analysis document challenges in specifying context and requirements, with users refining requests over successive prompts \citep{drosos2024rubberduck}. Some analysts may leave the choice of prediction target and estimation procedure to the LLM. Some analysts may describe the setting, explaining how predictions will inform decisions. Other analysts will provide explicit methodological guidance, including what to predict, the assumptions needed to estimate it, and/or how to construct the predictor. Prior work finds that LLM behavior depends on whether requirements are explicitly stated \citep{yang2025underspecification}, and StatQA evaluates whether providing statistical methods' applicability conditions helps the selection of relevant variables and appropriate methods \citep{zhu2024statqa}. We examine whether LLMs address decision dependency without this guidance and which forms of prompt information support target-aligned implementation.

\begin{figure}[t]
\centering
\includegraphics[width=0.88\linewidth]{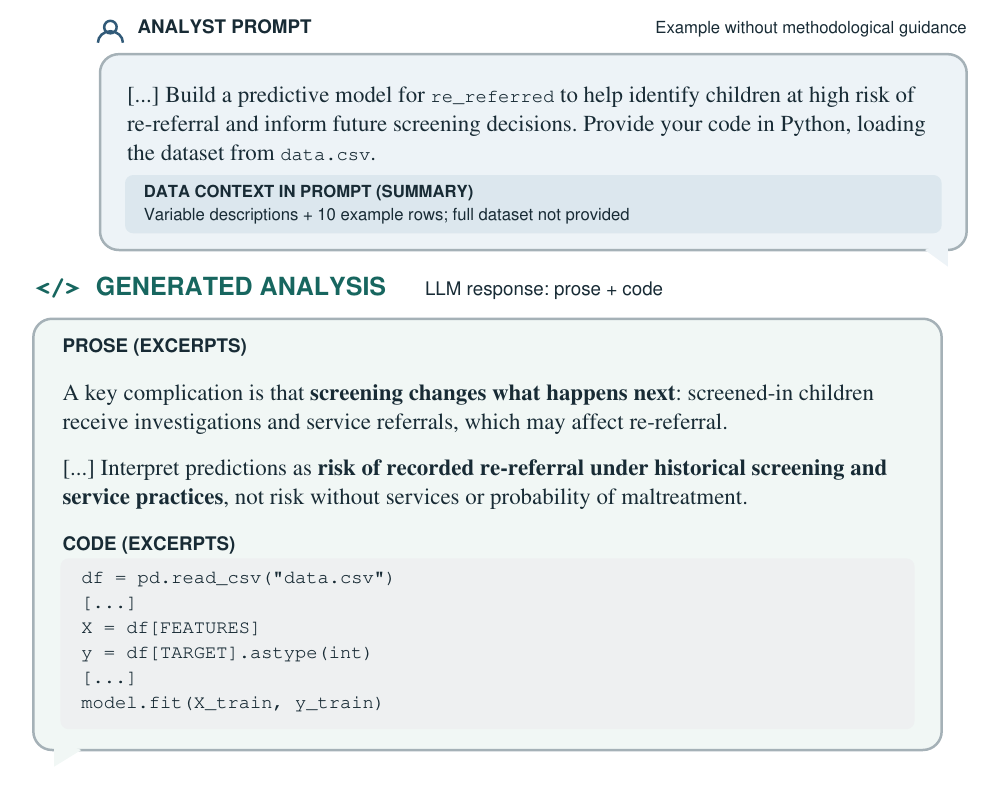}
\caption{Example prompt and generated analysis in child welfare
(\texttt{gpt6astra}, no-hint condition). The response acknowledges that
screening affects outcomes and explicitly describes its predictions as risk
under historical screening practices, not risk without services. Prose and
code are excerpted; ellipses mark omissions. The full prompt also provides
a data dictionary and ten example rows from the
dataset. }
\Description{Two dialogue bubbles show an analyst's request and an LLM response.
The response is labeled generated analysis and contains prose and code excerpts.
Its prose acknowledges intervention effects and disclaims predicting untreated risk.}
\label{fig:generated-analysis}
\end{figure}

We refer to the code and accompanying prose in an LLM response as the
\emph{generated analysis}. Analysts use natural-language explanations, alongside code and data artifacts, to understand and verify AI-generated analyses \citep{gu2024analysts}. However, even when an LLM verbally acknowledges decision dependency, this does not guarantee that the accompanying code addresses decision dependency. We therefore examine both prose explanations and code implementations. 
Figure~\ref{fig:generated-analysis} illustrates the prompt and the two parts of a generated analysis.

We evaluate generated code and explanations
in child-welfare, healthcare, and employment settings. When the code implements a procedure that estimates our target, we say it is \emph{target-aligned}. Across seven prompt
conditions, we vary the contextual and methodological guidance supplied to
11 models from two vendors. After evaluating six models in August 2026, we extended the study to five newer models available in September to assess whether newer models were more likely to address decision dependency. We refer to these two sets of models as the \emph{initial model set} and \emph{frontier extension set}, respectively. We investigate the following research questions:
\begin{enumerate}
\item[\textbf{RQ1.}] Do LLMs generate predictive modeling code that addresses
decision dependency without explicit methodological guidance? (no prompt hint)
\item[\textbf{RQ2.}] What prompt information is sufficient for LLMs to address decision dependency in the generated code?

\item[\textbf{RQ3.}] How does acknowledgment of decision dependency in explanations relate to target-aligned implementation?
\item[\textbf{RQ4.}] Are newer frontier models more likely to address decision dependency than earlier-generation models?
\end{enumerate}

We find that LLMs often fail to address decision dependencies without guidance in the prompt. For the child-welfare and healthcare settings, none of the models in the initial model set addressed decision dependencies without prompt guidance (Table~\ref{tab:cross-domain}, first row). 
In the frontier extension set, 2\% of the healthcare responses addressed decision dependency, as did 23\% of child-welfare responses  
(Table~\ref{tab:cross-domain}, extension no\_hint). We observed much higher rates of target-aligned implementation for the employment setting (20\% in the initial set and  78\% in the frontier extension set). Prompts specifying the target,
identifying assumption, and estimation procedure yield 99--100\%
target-aligned implementation. Less prescriptive prompts describing the
intervention and intended use also produce nearly universal target-aligned implementation in the frontier extension set, while results
are mixed in the initial model set.

Notably, LLMs may verbally acknowledge decision dependency
but fail to adequately address it in the code. For the no hint prompt condition, we observed this mismatch in 57\% of child-welfare responses (86/150), 40\% of healthcare responses (60/150), and 14\% of employment responses (21/150) from the frontier extension set. Under the three context prompts, the mismatch was uncommon in the frontier extension set but persisted in the initial model set, occurring in 8.1\% of child-welfare responses (44/540), 5.4\% of healthcare responses (29/540), and 2.6\% of employment responses (14/540). Some responses explicitly noted that their code predicted outcomes under historical decisions rather than outcomes without intervention. %

We draw on these findings to propose guidelines for prompt formulation and
reviewing generated analyses. These guidelines were informed by our study of decision dependency, but we hope they may be instructive in addressing other validity challenges in problem formulation. However, further study is needed to assess whether the  guidelines improve practice.

Our study investigates one aspect of problem formulation among many open questions and challenges around the use of LLMs to support predictive modeling in high-stakes decisions. 
Further inquiry is needed to investigate other validity challenges, such as proxy outcomes. 
Moreover, validity does not establish that a predictive system should be deployed in settings like child welfare, where concerns about surveillance, institutional
power, and systemic harms require careful consideration
\citep{eubanks2018automating, saxena2020human, saxena2023risknlp, stapleton2022imagining,wang2023against, kawakami2024studying}. 

\section{Related Work}
\label{sec:related-work}

\subsection{Problem formulation and measurement error}

Problem formulation determines how a practical goal becomes a question that
can be answered with data. Through an ethnographic study of data science
work, \citet{passi2019problem} show that selecting targets and translating
organizational goals into technical objectives involve judgment and
negotiation. \citet{jacobs2021measurement}
explain how observable measures can fail to represent the concepts they are
intended to capture, and they argue that making measurement assumptions explicit is important to
identify such failures. Both papers show how choices of prediction targets and their measurement are critical for validity and fairness.

\citet{coston2022validity} provide a validity framework for assessing whether predictive algorithms and their evaluations measure the quantities they are intended to target. Decision
dependency is one challenge within this broader framework.
\citet{coston2020counterfactual} show why risk under historical decisions can
differ from risk under a specified alternative decision, and they develop methods
for counterfactual evaluation. Our study asks whether LLM-generated analyses
address this distinction in their code and explanations.

A common theme in both predictive modeling and LLM evaluation is whether the measured outcome represents what one actually cares about. Research on LLM evaluation explores whether evidence from constructed benchmarks actually measures the capabilities they are intended to evaluate
\citep{salaudeen2025measurement,bean2025construct}. Our focus is whether an LLM-generated analysis
estimates the outcome needed for the intended high-stakes decision.

\subsection{Predictive analytics in high-stakes settings}

Prior work examines how predictive risk assessments are developed, evaluated, and used in real high-stakes decision making. \citet{chouldechova2018afst} document
the development, validation, and deployment of a child-welfare screening
model. Through interviews and observations of child-welfare workers,
\citet{kawakami2022improving} find that use of algorithmic advice depends on
workers' contextual knowledge, organizational pressures, and perceived
mismatches between predictions and decision-making objectives. These studies motivate why a prediction model must be evaluated in relation to the task-relevant decisions and real-world workflows.

Recent work also examines LLMs as predictors and decision-support tools in
high-stakes settings. \citet{mallari2024generative} compare generative model predictions
against traditional model predictions and human predictions in a recidivism context, finding that LLMs generally underperform and that bias-mitigation prompts can have unintended effects on predictions. \citet{lee2026auditing} audit accuracy and group
disparities in housing-placement prediction using tabular records and
outreach casenotes, finding that fine-tuning with casenote summaries can
improve both accuracy and fairness metrics. In child welfare,
\citet{moon2026promises} investigate language tools for tracking case
progress, identifying opportunities to help workers find relevant information
and areas where professional judgment is still required.

Other research questions whether predictive systems function
as intended and whether they should be deployed in these high-stakes settings. \citet{raji2022fallacy}  argue that
AI is often presumed to have basic functionality, but in many cases may fail to meet this basic requirement. \citet{wang2023against} identify recurring harms and limitations of systems
that use predictions about people to make decisions about them, and argue
that developers bear the burden of justifying their use. Our experiments
assess whether LLM-generated procedures estimate an intended target. They do not
establish that deploying those procedures would benefit the people affected
or be justified in a particular context.

\subsection{LLM-assisted data science}

LLM-based systems already assist with several stages of data science, from writing analysis code to conducting machine-learning experiments. For example, CliMB lets clinician scientists develop models through
natural-language interaction, guiding them through a machine-learning
workflow \citep{saveliev2024climb}. Prior evaluations examine whether LLMs produce working code, build accurate predictive models, and formulate statistical problems appropriately. DS-1000 tests solutions to data-science coding
problems \citep{lai2023ds1000}; DSBench evaluates data analysis and modeling
on realistic datasets \citep{jing2025dsbench}; and MLAgentBench evaluates
agents carrying out machine-learning experiments
\citep{huang2024mlagentbench}. DataSciBench evaluates data-science outputs using executable criteria tied to the task's stated goals \citep{zhang2026datascibench}. LLM4DS-Benchmark evaluates data-science code
generation across correctness, consistency, feedback-guided recovery, and
cost \citep{boominathan2026empirical}. StatLLM provides statistical
analysis tasks and human-written SAS solutions, alongside LLM-generated solutions with expert evaluations of their quality, executability, and output accuracy \citep{song2026statllm}.

Other benchmarks explicitly examine analytical choices. BLADE compares
agents' variable choices, transformations, and statistical models with
expert analyses of open-ended research questions \citep{gu2024blade}.
 StatFormBench examines whether LLMs can classify statistical task categories and identify relevant variables \citep{wang2026benchmarking}. We investigate a complementary aspect of problem formulation: whether the procedures generated by LLMs estimate a target appropriate to the intended decision. Specifically, we test whether generated code accounts for how historical decisions affected recorded outcomes and whether accompanying explanations accurately describe the procedure and its assumptions. 

Prior work has also examined how prompt specification affects
performance. \citet{yang2025underspecification} find that LLMs can infer
unstated requirements, but their behavior on those requirements is less
stable across model and prompt changes. \citet{zhu2024statqa} also examine whether explaining when statistical methods can be used helps LLMs select appropriate statistical methods. \citet{bertran2026many} find that
autonomous AI analysts produce divergent estimates and conclusions from the
same data, hypothesis, and prespecified estimand. Changing the underlying
model or prompted analyst persona shifts the distribution of conclusions,
including among analyses that pass their methodological audit. 
\citet{rewolinski2026sanity} test whether agents reach conclusions that remain supported when the data, metadata, or prompts are perturbed. This literature demonstrates that LLM-generated analyses are sensitive to how a request is specified. Therefore, we vary the contextual and methodological information in the prompt to examine when LLM-generated analyses address decision dependency.

Recent work suggests that LLM agents can influence analytical choices. 
In a randomized study, \citet{herdagdelens} find that expert analysts followed their LLM agents' randomly assigned preferences in 86\% of choices concerning outcome measures, comparison methods, target populations, and weighting. This finding highlights the importance of understanding whether LLMs can appropriately make design choices during problem formulation in high-stakes settings.

\section{Study Design}
To evaluate whether LLM-generated analyses address decision dependency, we ask LLMs to write predictive-modeling code and explain their approach in
child-welfare, healthcare, and employment settings. In each setting, historical
decisions affect the recorded outcomes. We investigate whether LLM-generated code accounts for these effects when predicting outcomes without intervention. 

\subsection{Target estimand}
\label{sec:target-estimand}
In many high-stakes decision-making settings, decision makers are interested in assessing the risk of adverse outcomes under no/minimal intervention.
For instance, in child-welfare call screening, workers are tasked with
identifying children at high risk of re-referral if screened out \cite{chouldechova2018afst, coston2020counterfactual}. Historical records include both children who were historically screened in and children who were historically screened out. Children who were screened in, and consequently received investigations and services, may have different outcomes than they would counterfactually have had if they had been screened out. Therefore, a model trained to predict
these historically recorded outcomes can misestimate risk without intervention
and misdirect screening decisions. 

For each individual, $X$ is a vector of observed features, $T \in \{0, 1\}$ indicates whether they received the intervention, and $Y$ denotes the recorded outcome. Predictors include binary indicators and numerical variables. Our outcome is binary in the child-welfare domain and continuous in the healthcare and employment domains. 

We use ``treated'' and ``untreated'' to refer to individuals with $T=1$
and $T=0$, respectively. Let $Y(0)$ denote the outcome an individual
would have without intervention. By consistency, $Y=Y(0)$ for untreated
individuals. For treated individuals, $Y(0)$ is unobserved.\footnote{We
assume consistency, meaning that the recorded outcome equals the
potential outcome under the intervention actually received. We also
assume no interference, meaning that one individual's outcome does not
depend on interventions received by others.}

For the modeling task, the intended target is $\mathbb{E}[Y(0)\mid X]$, the expected outcome without intervention given the observed features. This can differ from $\mathbb{E}[Y\mid X]$, the expected recorded outcome under historical decisions. In child welfare, for example, the latter estimand averages outcomes under both screen-in and screen-out according to historical screening decisions. 

To identify the target, we assume conditional ignorability, so $Y(0)$
and $T$ are independent given $X$, and overlap, so
$p_0(X)=\Pr(T=0\mid X)>0$ wherever predictions are needed. Under these
assumptions and consistency, two formulas identify the same target
from the observed data.

The outcome-regression formula follows by conditioning on untreated
cases.
\begin{equation}
\label{eq:target-regression}
\mathbb{E}[Y(0)\mid X]
= \mathbb{E}[Y(0)\mid X,T=0]
= \mathbb{E}[Y\mid X,T=0].
\end{equation}
Conditional ignorability gives the first equality, and consistency
among untreated cases gives the second.

The inverse-propensity-weighting formula uses weights
$w=\mathbf{1}\{T=0\}/p_0(X)$.
\begin{equation}
\label{eq:target-ipw}
\mathbb{E}[Y(0)\mid X]
= \mathbb{E}[wY(0)\mid X]
= \mathbb{E}[wY\mid X].
\end{equation}
Conditional ignorability and $\mathbb{E}[w\mid X]=1$ give the first
equality. Consistency gives the second because only untreated cases
have nonzero weight.

We use these formulas to explain which implementations the scoring
rubric accepts (\S\ref{sec:metrics}).

\subsection{Request structure}
Each request to the LLM provides a description of the modeling problem, a data dictionary,
and ten example rows, and asks for Python code that loads \texttt{data.csv}.
We vary the contextual and methodological guidance supplied in the request
and evaluate the resulting code and explanation. We use an LLM judge to score
the generated analyses against a rubric on various measures and check its labels
against an author's manual assessments. We evaluate the code and prose
without executing the code.

\subsection{Measures of generated responses}
\label{sec:metrics}

To assess whether LLMs generate code that addresses decision dependency
(RQ1), our primary measure, \emph{target-aligned implementation}, checks
whether the code uses an accepted procedure to estimate the expected
outcome without intervention.

The rubric explicitly accepts three implementations. Two estimate the
conditional mean in Equation~\eqref{eq:target-regression}: fitting a
model on untreated cases only, and fitting a model on both groups
with $T$ as a predictor and setting $T=0$ at prediction time. The third
fits a conditional mean using the inverse-propensity weights $w$ and
targets Equation~\eqref{eq:target-ipw}.

In addition to target-aligned implementation, we also assess four other measures: whether the response verbally acknowledges decision dependency, accurately describes
its code, states assumptions linking its procedure to an estimand, and claims to estimate the intended target.
Table~\ref{tab:response-measures} defines these five measures. To answer RQ3, we compare verbal acknowledgment with target alignment in responses, to see if code and prose are aligned in addressing decision dependency.

\begin{table}[!ht]
\centering
\small
\setlength{\tabcolsep}{4pt}
\renewcommand{\arraystretch}{1.15}
\begin{tabular}{@{} >{\raggedright\arraybackslash}p{0.24\linewidth}
                  >{\raggedright\arraybackslash}p{0.46\linewidth}
                  >{\raggedright\arraybackslash}p{0.25\linewidth} @{}}
\toprule
Measure & Question assessed & Response scale \\
\midrule
Target-aligned implementation & Does the code use an accepted procedure
for estimating the outcome without intervention? & Binary: 0 or 1 \\
\addlinespace
Verbal acknowledgment & Does the explanation connect intervention effects
to the problem with training on recorded outcomes? & Binary: 0 or 1 \\
\addlinespace
Faithful description & Does the explanation accurately describe
what the code implements? & Binary: 0 or 1 \\
\addlinespace
Assumption reporting & Does the response name an assumption
linking its procedure to an estimand? & None; named but vague; named and specific \\
\addlinespace
Claim about the target & Does the response assert, qualify, or disclaim that
its output represents the intended target? & Assertive; hedged; disclaimed \\
\bottomrule
\end{tabular}
\caption{Five measures used to evaluate generated code and explanations. Appendix~\ref{app:metric-details} details how each measure is scored as well as which responses are included when calculating rates.}
\label{tab:response-measures}
\end{table}

Estimating the target outcome from observed data requires identifying assumptions \citep{lundberg2021estimand}. We therefore check whether
responses name an assumption required by the procedure they choose. We call this \emph{assumption reporting}. 
We calculate assumption-reporting rates only for LLM responses that use procedures that are covered by this measure (Appendix~\ref{app:as}).

We record whether the response asserts
that its analysis estimates the intended outcome, qualifies its assertion,
or explicitly says it does not. Responses that do not make a statement about whether their analysis targets the intended outcome are
grouped with qualified claims. We call this measure a \emph{claim about the
target}. Appendix~\ref{app:sc} gives the classification rules. We use this measure to identify responses that assert they estimate the intended target despite problems in their implementation or explanation (Section~\ref{sec:results-misleading}). Appendix~\ref{app:elicited} reports a separate exploratory follow-up on elicited confidence, in which we asked LLMs how confident they were that their approach addressed decision dependency.

\paragraph{Misleading claims about the target.}
Using the judge's labels, we classify a response as making a misleading claim
when it asserts that its
analysis estimates the intended outcome (claim about the target = assertive), but its code uses
a procedure that does not implement the intended target (target-aligned
implementation = 0) or its explanation inaccurately describes the
code (faithful description = 0). Section~\ref{sec:results-misleading} reports how frequently this occurs.

\subsection{Varying prompt informativeness}
\label{sec:prompt-informativeness}

In practice, analysts may provide different amounts and kinds of information
when requesting a predictive model. We use the no-hint condition to assess
whether LLMs address decision dependency without explicit methodological
guidance (RQ1). For RQ2, we compare responses from different prompts to assess what information ultimately helps LLMs produce code that targets the intended outcome. We construct prompts to vary dimensions of problem formulation identified by \citet{lundberg2021estimand}: defining what quantity to estimate, stating the assumptions that make it identifiable from observed data, and actually estimating it from data.
We vary which of these elements the prompt supplies, along with practical
information about the decision setting. 

Through seven prompt conditions, we vary the methodological guidance while holding the rest of the prompt fixed within
each domain and dataset version; these conditions range from an implicit target to a fully specified analysis. The \texttt{no\_hint} condition provides the general domain context and general description of data without explicitly stating the intended target. This condition tests whether LLMs infer the intended target from the provided base context. Three context prompts add information in plain language about how intervention affects outcomes and when predictions
will be used: \texttt{context} leaves the outcome without
intervention implicit, while \texttt{context\_target} states it minimally
and \texttt{context\_target\_detail} provides a fuller definition.
The \texttt{estimand\_assumption} condition expresses the target in the potential-outcomes
notation introduced in Section~\ref{sec:target-estimand} and instructs the LLM to assume conditional ignorability: given the observed features,
intervention assignment is independent of potential outcomes.
The \texttt{procedure} condition instructs the model to train on untreated
cases and explains why treated outcomes do not directly represent the target.
Finally, \texttt{full\_specification} supplies the formal target, assumption,
and estimation procedure together; this condition can be seen as a combination of the \texttt{estimand\_assumption} and \texttt{procedure} conditions. Appendix~\ref{app:prompts} provides prompt wording.

\subsection{Domains}
\label{sec:domains}

\paragraph{Child welfare.} Child-welfare hotline workers decide whether to screen
in a referral about a child for investigation and possible services, with the goal of
prioritizing children at high risk of harm
\citep{chouldechova2018afst, coston2020counterfactual}.
In particular, hotline workers must assess the risk of harm if they do not intervene -- that is, if they screen the case out of investigation. We denote the binary screening decision as $T$, where $T=1$ indicates screening in to investigation.
Because harm is not directly
observed, we let the outcome $Y$ be re-referral within six months. The target is the risk of re-referral if screened out, which may differ from observed re-referral outcomes for screened in cases since investigation and services may change subsequent re-referral risk. 

We use synthetic hotline
screening records based on the child welfare screening context of the Allegheny County Family Screening Tool \cite{chouldechova2018afst, coston2020counterfactual}. We generate synthetic child-welfare records with two underlying risk
structures, described in Appendix~\ref{app:dgp}. 
We construct the data to reflect the screening rate and average
re-referral risk under screen-out in the Allegheny County dataset used
by \citet{coston2020counterfactual}.
Appendix~\ref{app:dgp} reports these quantities and compares the
synthetic data against the real-data.
No real
child-welfare records were used in the prompts. 

\paragraph{Healthcare.} Health systems enroll patients in care-management programs,
which provide additional resources to patients with complex health needs. These care-management programs often aim to
prioritize patients predicted to have the greatest future need
\citep{obermeyer2019dissecting}. Therefore, the decision $T$ is program enrollment, and the
outcome $Y$ is healthcare cost. We note that cost is an imperfect proxy for health need
\citep{obermeyer2019dissecting}. Because enrollment may change a patient's subsequent healthcare
cost, the intended target is expected cost without enrollment.

We use the synthetic patient-year dataset from
\citet{obermeyer2019dissecting}, which they generated using \texttt{synthpop}.
The intervention is enrollment in a care-management program (approximately
0.9\% of records). Because this dataset comes from a published study, models may
have encountered its source during training. We modify the data
to reduce recognition and check whether responses identify the
source (\S\ref{sec:disguise}).

\paragraph{JOBS (employment).} Public employment services use statistical profiling to direct support
toward jobseekers at risk of long-term unemployment
\citep{desiere2019profiling}. Targeting can prioritize those with the
poorest prospects without assistance or those expected to benefit most
from it \citep{lechner2007caseworkers}. In our study, we suppose the goal is to serve those with the lowest expected
earnings without the program. The decision $T$ is enrollment in a job-training
program, the outcome $Y$ is subsequent earnings, and the target is expected earnings
without enrollment. 

We use data from the National Supported Work
job-training demonstration and a non-experimental comparison group
\citep{lalonde1986evaluating}, starting from the same sample composition as
\citet{guerdan2023counterfactual}. The intervention is program enrollment
(approximately 9.9\% of records). Unlike
\citet{guerdan2023counterfactual}, who study unemployment, we predict earnings. As with healthcare, we modify these published data to reduce
source recognition and check whether responses identify their
origins (\S\ref{sec:disguise}).

\begin{table}[t]
\centering
\small
\setlength{\tabcolsep}{4pt}
\renewcommand{\arraystretch}{1.15}
\begin{tabular}{@{}
  >{\raggedright\arraybackslash}p{0.12\linewidth}
  >{\raggedright\arraybackslash}p{0.27\linewidth}
  >{\raggedright\arraybackslash}p{0.25\linewidth}
  >{\raggedright\arraybackslash}p{0.29\linewidth}
@{}}
\toprule
Domain & Decision-making goal & Intended prediction target & Data source \\
\midrule
Child welfare
& Prioritize children at high risk if screened out
& Re-referral risk within six months under screen-out
& Synthetic records constructed for this study, informed by
  \citet{coston2020counterfactual} \\
\addlinespace
Healthcare
& Prioritize patients with greater health need, using cost as a proxy
& Expected healthcare cost without care-management enrollment
& Synthetic patient-year data from
  \citet{obermeyer2019dissecting}, modified for this study \\
\addlinespace
Employment
& Prioritize people with the lowest expected earnings without training
& Expected earnings without job-training enrollment
& National Supported Work and PSID comparison data
  \citep{lalonde1986evaluating,dehejia1999causal},
  modified for this study \\
\bottomrule
\end{tabular}
\caption{We investigate whether LLMs address decision dependency in three domains: child welfare, healthcare and employment. In each setting, the prediction target is an expected outcome without intervention,
$\mathbb{E}[Y(0)\mid X]$. We note that healthcare cost is an imperfect proxy
for health need.}
\label{tab:domain-overview}
\end{table}

\subsection{Model sets and response collection}
\label{sec:models}
We evaluate 11 models from OpenAI and Anthropic in two sets. The \emph{initial model set}
consists of \texttt{haiku}~4.5, \texttt{sonnet}~4.6, \texttt{opus}~4.8,
\texttt{gpt5nano}, \texttt{gpt5}, and \texttt{gpt5sol}, evaluated in
August 2026. The \emph{frontier extension set} comprises newer models available in September 2026: \texttt{gpt6astra}, \texttt{gpt6sol},
\texttt{fable5.1}, \texttt{opus5.5}, and \texttt{sonnet5.5}. Appendix~\ref{app:models} lists model identifiers and collection dates.
To answer RQ4, we compare the initial and frontier extension sets under each prompt condition and domain to determine whether newer frontier models handle decision dependency more reliably.

Repeated requests let us examine variation in generated responses at a fixed
prompt, an issue highlighted by research on reproducible LLM evaluation
\citep{blackwell2025reproducible}.
Each model receives 30 independent requests for each of seven prompt
conditions in each dataset. This yields 840 responses per model and 9,240 in total. We report the initial and frontier extension sets separately to show
differences between the tested sets.

\subsection{Reducing and checking dataset recognition}
\label{sec:disguise}

Prior work documents LLM memorization of public tabular datasets
and investigates how data transformations affect model performance
\citep{bordt2024elephants}.
Since the healthcare and JOBS datasets derive from published sources, LLMs may have encountered the data or descriptions of the original studies during their training. Before constructing the prompts, we renamed variables, slightly jittered numeric values, and removed identifying columns to reduce dataset recognition (Appendix~\ref{app:disguise}). After modifying the datasets, we checked whether LLM responses still identified data sources. For the initial model set, we read responses generated for this check and looked for source names or references to findings from the original studies. For the frontier extension set, we searched responses for source
names. Appendix~\ref{app:audit} describes these checks.

\subsection{Automated scoring and author validation}
\label{sec:judge-validation}

We use an LLM judge to score the measures defined in Section~\ref{sec:metrics} against an explicit, multidimensional rubric, and validate its scores against author judgments (initial validation: $n=252$; frontier validation: $n=171$). %
The scoring model that we use as the judge, \path{claude-opus-4-8}, and final rubric are held fixed across domains and both model sets. It scores verbal
acknowledgment, code implementation, and faithful description together, then
scores target claims and assumption reporting in separate calls. It scores assumption reporting based on the estimation procedure identified in the first call (scoring code implementation).

An author manually scored validation samples using the same rubric. Following prior work \citep{bavaresco2025judges}, we
compared judge labels against author labels on different combinations of domains,
models, and measures. For the initial model set, all 20 tested combinations met our agreement benchmark of Cohen's
$\kappa\geq0.8$, which adjusts for agreement expected by chance
\citep{landis1977measurement}. In the frontier core validation sample,
target-aligned implementation labels agreed with author judgments in
104 of 105 responses. On an additional
sample of 21 Sonnet 5.5 responses, judge and author labels for target-aligned
implementation agreed in all 21 cases.
Appendices~\ref{app:judge-val} and~\ref{app:judge-frontier} describe the validation samples and report agreement with author judgments.

\section{Results}
\label{sec:results}

Our findings indicate that in many settings LLMs fail to address decision dependency.
In the initial model set, no responses correctly addressed decision dependency in the child-welfare or healthcare settings without guidance. In contrast, prescriptive prompts specifying the estimand, identifying assumption, and estimation procedure yielded 99--100\% target-aligned implementation across model vendors and datasets. Models in the frontier extension set were substantially more likely than those in the initial set to address decision dependency under prompts providing intermediate levels of guidance. Finally, we find that LLMs sometimes acknowledge decision dependency in their explanations without addressing it in their generated code.

We report the initial model set and the frontier extension set separately, comparing performance across the
child-welfare, healthcare, and JOBS datasets. When calculating target-alignment rates, we exclude responses that do not
produce a method for predicting outcomes for individual people---for example,
when an LLM declines to build the requested model.
Appendix~\ref{app:main-denominators} describes response exclusions.

\subsection{LLMs struggle to address decision dependency without methodological guidance (RQ1)}
\label{sec:results-no-hint}

Without methodological guidance (\texttt{no\_hint}), target-aligned
implementation is 0\% in child welfare (0/169) and healthcare (0/180), and
20\% in employment (36/180) for the initial model set. In the frontier
extension set, target-aligned implementation increases to 23\% in child welfare
(34/150), 2\% in healthcare (3/150), and 78\% in employment (117/150)
(Table~\ref{tab:cross-domain}). Two models in the frontier extension set, \texttt{fable5.1} and
\texttt{opus5.5}, produce target-aligned code in all 30 employment responses, but
only 7/30 and 24/30 child-welfare responses, respectively.
Of the three datasets, the employment dataset (JOBS) is the most well-known and is widely studied in the causal inference literature, where intervention effects and decision dependencies are carefully addressed. We suspect that models performed best on the JOBS dataset because models may have been exposed to this literature during training. Full per-model results are in Appendices~\ref{app:tables-original}
and~\ref{app:new-results}.

\subsection{Specifying target, assumptions, and procedure achieves target-aligned implementation (RQ2)}
\label{sec:results-ladder}

 Prompts that specify the target, identifying
assumption, and procedure (\texttt{full\_specification}) achieve rates of target alignment of
99--100\% and 100\% in the initial model set and frontier extension set, respectively. In the frontier extension set, prompts that provide any additional information about the context, the target, and learning procedure (all prompts except \texttt{no\_hint}) address decision dependency at least 99\% of the time: 
Across the three context prompts (\texttt{context},
\texttt{context\_target}, and \texttt{context\_target\_detail}) and all
three domain settings, the frontier extension set produces target-aligned
code in 1,345/1,350 responses. This total
includes 448/450 responses under the \texttt{context} prompt, which leaves
the target implicit. For these particular datasets and newer models, describing the
practical setting appears to be sufficient for eliciting the intended analysis without
formal causal notation or explicit specification of the estimation procedure. In contrast, models in the initial set had
target-aligned implementation under the \texttt{context} prompt at substantially lower rates: 51\% in child welfare, 57\% in healthcare,
and 51\% in employment 
(RQ4; Table~\ref{tab:cross-domain}). 

\begin{table}[!ht]
\centering
\small
\begin{tabular}{@{} ll ccc ccc @{}}
\toprule
& & \multicolumn{3}{c}{Verbal acknowledgment rate} & \multicolumn{3}{c}{Target-align. implem. rate} \\
Model set & Condition & Child welfare & JOBS & Health & Child welfare & JOBS & Health \\
\midrule
Initial & no\_hint & 0.17 & 0.16 & 0.04 & 0.00 & 0.20 & 0.00 \\
Initial & context & 0.54 & 0.52 & 0.58 & 0.51 & 0.51 & 0.57 \\
Initial & context\_target & 0.58 & 0.71 & 0.67 & 0.56 & 0.79 & 0.67 \\
Initial & context\_target\_detail & 0.67 & 0.72 & 0.69 & 0.68 & 0.78 & 0.72 \\
Initial & estimand\_assumption & 0.94 & 0.97 & 0.95 & 0.93 & 0.99 & 0.96 \\
Initial & procedure & 0.51 & 0.61 & 0.56 & 0.99 & 1.00 & 1.00 \\
Initial & full\_specification & 0.90 & 0.95 & 0.88 & 0.99 & 0.99 & 1.00 \\
\midrule
Extension & no\_hint & 0.80 & 0.81 & 0.42 & 0.23 & 0.78 & 0.02 \\
Extension & context & 1.00 & 0.97 & 0.97 & 1.00 & 0.99 & 0.99 \\
Extension & context\_target & 1.00 & 0.97 & 0.99 & 0.99 & 1.00 & 0.99 \\
Extension & context\_target\_detail & 1.00 & 1.00 & 1.00 & 1.00 & 1.00 & 0.99 \\
Extension & estimand\_assumption & 1.00 & 1.00 & 1.00 & 1.00 & 1.00 & 1.00 \\
Extension & procedure & 0.72 & 0.84 & 0.68 & 1.00 & 1.00 & 1.00 \\
Extension & full\_specification & 1.00 & 1.00 & 1.00 & 1.00 & 1.00 & 1.00 \\
\bottomrule
\end{tabular}
\caption{Verbal acknowledgment and target alignment. The three leftmost columns report rates of verbal acknowledgment; the three rightmost columns report rates of target-aligned implementation. Initial-set
denominators are 169--180 in child welfare and 180 in JOBS/Health; extension denominators are
150 throughout.}
\label{tab:cross-domain}
\end{table}

\subsection{Verbal acknowledgment does not guarantee target-aligned code (RQ3)}
\label{sec:results-harm}

An LLM response verbally acknowledging decision dependency does not
guarantee target-aligned code. Without methodological guidance in the prompt,
responses from the frontier extension set verbally acknowledge the problem but do not
address decision dependencies in the code in 57\% of child-welfare responses (86/150), 40\% of
healthcare responses (60/150), and 14\% of employment responses (21/150). Under the three context prompts, we observed acknowledgment
without target-aligned implementation in 8.1\% of child-welfare responses (44/540),
5.4\% of healthcare responses (29/540), and 2.6\% of employment responses (14/540) from the
initial model set, compared with 0.2\% (1/450), 0.7\% (3/450), and 0.2\% (1/450), respectively,
from the frontier extension set (RQ4).

For example, one employment response trains on recorded earnings from both
enrolled and non-enrolled people. Its explanation states: ``This model
predicts outcomes under the historical mix of enrollment and non-enrollment.
It does not estimate earnings without training or identify who would benefit
most from training.'' The response therefore acknowledges the difference
between its predictions and the intended target without claiming to have
addressed it.

Some of these responses explicitly state that their code does not estimate
outcomes without intervention. Among the frontier \texttt{no\_hint} responses
that acknowledge decision dependency but lack target-aligned implementation,
the judge identified such warnings in 63/86 child-welfare responses,
25/60 healthcare responses, and 21/21 employment responses.

\subsection{LLMs can make misleading claims about implementation}
\label{sec:results-misleading}

We next examine responses that claim to estimate the intended target despite problems in their code or explanation. Following Section~\ref{sec:metrics}, we classify a claim as misleading when the response asserts that it estimates the intended target, but its code does not implement that target or its explanation inaccurately describes the code. 
 For example, a child-welfare response under
\texttt{context\_target\_detail} provides an implementation that trains on recorded re-referral outcomes from \emph{both}
screened-in and screened-out cases. This model predicts risk under observed decisions, not risk under \emph{screen-out}. Yet the response misleadingly claims: ``The final model allows you to rank
cases by predicted risk of re-referral if a case were not screened in.''

In the initial model set, misleading claims were most frequent under the three \texttt{context} prompts, occurring in 8.1\% of responses across these prompts (131/1,619), compared with 3.8\% without methodological guidance (20/529). In the frontier extension set, misleading claims occurred in 0.15\% of responses under the context prompts (2/1,350) and 1.1\% without methodological guidance (5/450).
In the initial model set, 129 of the
131 misleading responses claim to have estimated the intended target, but in reality the code does not implement a target-aligned procedure. The remaining two responses assert that they estimate the intended target and have target-aligned code, but their explanations inaccurately describe that code. Per-model counts across all seven prompt conditions
are reported in Appendices~\ref{app:tables-original} and~\ref{app:new-results}. Appendices~\ref{app:judge-val} and~\ref{app:judge-frontier} report author
validation of the underlying labels.

\subsection{LLMs often fail to state key assumptions}
\label{sec:results-as}

The implementation procedure we classify as target-aligned is, in reality, only target-aligned when key assumptions hold. These assumptions include conditional ignorability and overlap
(\S\S\ref{sec:metrics} and~\ref{sec:prompt-informativeness}).
An LLM response should state these assumptions,
and the analyst should assess whether those assumptions are appropriate
for their setting.

Under the \texttt{procedure} prompt, which instructs the model to train only on untreated cases, the responses named at least one assumption in a mere 2.4\% (13/534)
of responses with target-aligned implementations from the initial model
set and 21.1\% (95/450) from the frontier extension set. The models were most likely to name an assumption for the child-welfare setting: 4.0\% (7/174) for the initial set
and 36.0\% (54/150) for the frontier extension set. For the healthcare setting, the rates were 2.8\%
(5/180) for the initial set and 11.3\% (17/150) for the frontier extension set. For the employment setting,  the rates were 0.6\% (1/180) for the initial set and 16.0\% (24/150)
for the frontier extension set. Of the 54 frontier child-welfare responses that
name an assumption, 29 are from \texttt{opus5.5} (29/30 responses).
Appendices~\ref{app:judge-val} and~\ref{app:judge-frontier} report author
validation of this measure.

\section{Implications}
\label{sec:discussion}

Our findings highlight the importance of providing informative prompts that specify the intended prediction target and of validating whether the generated code is aligned with the target. In this section, we translate our findings into actionable guidelines for prompt formulation and reviewing LLM-generated analyses.  While grounded in our empirical findings, these guidelines have not been evaluated in real-world workflows. 
These guidelines were motivated by predictive modeling, but we anticipate that they may be more broadly applicable to statistical and computational tasks that estimate a target quantity. Future work should evaluate the applicability of these guidelines to these other settings and their effectiveness in real-world practice.

\subsection{Model prompt guidelines}

\paragraph{Explain what the model should predict (theoretical estimand) and how we can identify this in the observed data (empirical estimand).}
Specifying the prediction task may appear straightforward, but proper specification requires careful consideration and sufficient detail. 
For instance, although the \texttt{no\_hint} condition instructed models to predict ``re-referred,'' this was largely insufficient to ensure alignment with the intended prediction target (\S\ref{sec:results-no-hint}).
The request should describe the intended target and how it relates to
historical decisions. 
We recommend beginning with the \emph{theoretical estimand}, which is a hypothetical quantity stating the substantive or scientific target \citep{lundberg2021estimand}.
In our child-welfare example, the theoretical estimand is a child's
risk of harm if screened out.

\paragraph{Explain how to identify the theoretical estimand in the observed data (empirical estimand).}
The request should specify assumptions that identify the theoretical estimand in the observed data as an \emph{empirical estimand}
\citep{lundberg2021estimand}. The request should use mathematical notation to express this empirical estimand.\footnote{
Although frontier models often generated target-aligned code under prompts that only used ordinary language to describe the setting (the three \texttt{context} prompts; \S\ref{sec:results-ladder}), we nonetheless recommend precise specification using both plain English and mathematical notation to reduce ambiguity.}
For example, if we assume that re-referral is a reasonable proxy outcome for risk of harm and additionally assume conditional ignorability and overlap, we can identify the theoretical estimand of a child's
risk of harm if screened out as $E[Y \mid X, T= 0]$.

\paragraph{Specify the outcome, predictive, and decision variables, and describe decision dependencies.}
In many high-stakes settings, particularly where the target is not directly observable, there may be multiple proxy outcomes available in the data \cite{obermeyer2019dissecting, mhasawade2024causalbias, guerdan2023groundless, watson2023multitarget}. %
The prompt should explicitly identify the  observed outcome variable $Y$. The prompt should also identify the historical decision $T$ and predictors $X$ available before the decision. In child welfare, \texttt{re\_referred} is $Y$ and \texttt{screened\_in} is $T$;
\texttt{prior\_referrals} is one pre-decision predictor. The prompt should state which predictors are available before
the decision and when the outcome is measured. The prompt should
also explain how variables are related to one another and, in the case of decision dependency, how decisions can affect outcomes. For example, screening a child in can lead to services that affect the child's subsequent risk of re-referral. 

\paragraph{Specify the training population.}
In the prompt, the analyst should specify which records to use for training and explain why those records are appropriate for estimating the intended outcome, rather than assume that the LLM will properly infer the correct population. In our study, prompting LLMs to train on untreated cases resulted in target-aligned code in almost all responses. However, many of these responses omitted the assumptions required by that approach (\S\ref{sec:results-as}). Therefore, it is important for analysts to state those assumptions in the prompt and assess whether they are plausible in their setting.

\subsection{Evaluation guideline}
\label{sec:checklist}

\paragraph{Specify an evaluation checklist.}
An LLM response's explanation can help analysts review its code, but the accompanying explanation is not always reliable. In our study, some responses claimed to estimate the intended target despite not implementing a valid procedure to do so (\S\ref{sec:results-misleading}).
Therefore, analysts should check what outcome the response claims to predict and the actual procedure implemented in its code. Concretely, this means checking which columns the code uses for $Y$, $X$, and $T$, which records it trains on, how it weights them, and how it generates predictions. The checklist should also specify possible issues pertaining to the population, data collection or changes in policy that could make the assumptions implausible \cite{coston2022validity}. 

These checks are intended to assess the validity of the procedure, not the accuracy of the predictive model. It is possible that a model may have high predictive accuracy but fail these checks, if for instance it used the wrong training population \cite{subbaswamy2020development, coston2022validity}. Our results show the importance of independently verifying these checks even when a response explicitly claims to estimate the intended target (Section~\ref{sec:results-misleading}). 
More broadly, verifying that the implementation is target-aligned does not establish
that a resulting predictive system would function
effectively in practice or provide meaningful benefits
\citep{raji2022fallacy}.
\looseness=-1

\paragraph{Formulate a plan for assessing broader ethical and sociotechnical concerns.}
In addition to evaluating the validity of the LLM-generated response, it is critical to have a deliberate plan for identifying, assessing, and addressing ethical and sociotechnical concerns including possible harms of deployment, unintended uses, societal impacts, and legal considerations. For an example of how to approach this, we refer readers to \citet{kawakami2024situate}. 
We observed in our study that the LLM response occasionally flagged ethical concerns and refused to perform the modeling task (44 child-welfare responses from the initial model set, 43 from \texttt{gpt5sol} and one from \texttt{gpt5}; Appendix~\ref{app:main-denominators}). We recommend that practitioners take seriously any concerns raised in the LLM response, but it is not sufficient to rely on the LLM to raise these concerns. Analysts must conduct careful deliberations with key stakeholders to identify and reason about these issues and impacts \cite{kawakami2024situate}.

\section{Limitations}
\label{sec:limitations}
This section considers the limitations of our study design and findings. First,
our study explores whether LLMs address a particular validity issue, decision dependency, but validity is a broad notion that encompasses many aspects not assessed in our study. Future work could investigate other validity-related challenges, including measurement error due to proxy outcomes, misalignment between the study and target populations \citep{coston2022validity}, and train-test leakage \citep{chouldechova2018afst}. These validity challenges can coexist. For example, in our healthcare setting, accounting for decision dependency does not resolve using healthcare cost as a proxy for health need.

Second, we evaluate LLM responses using an automated judge whose labels do not always perfectly agree with author judgments. Validation showed high agreement
for target-aligned implementation, but lower agreement for some
explanation measures (Appendices~\ref{app:judge-val}
and~\ref{app:judge-frontier}). Additional human annotations could help strengthen judge reliability and refine rubric criteria. Evaluating the same responses with different LLM judges could also reveal how sensitive our findings are to the choice of scoring model.

Third, the experimental settings may limit what we can conclude about
performance in practice. The child-welfare and healthcare settings
use synthetic records (\S\ref{sec:domains}), and we have not
established whether LLMs would respond similarly with real agency or clinical records. Some LLM responses named source datasets or related research despite our modifications to the data and
prompts (Appendix~\ref{app:audit}). We cannot determine how much
prior familiarity influenced the generated analyses or
differences in performance across settings.

Finally, we evaluate generated code and explanations without
deploying an LLM-assisted system. We did not measure how these
responses affected practitioners' beliefs about the analysis, their
reliance on it, or their subsequent decisions. For instance, we classify responses as misleading when the explanation describes a different implementation than what the code implements, but we do not establish that an
analyst was misled in practice by this mismatch. We also have not tested
how practitioners or institutions would use the proposed guidelines
in Section~\ref{sec:checklist}. The guidelines identify issues to explore, but their feasibility and effectiveness in practice should be assessed in future work. Importantly and as discussed earlier, the decision whether to deploy a predictive system in high-stakes decision making requires deliberating on considerations beyond those in focus for this paper, including the potential for misuse, institutional context, impacted communities, and potential harms.

\section{Conclusion}

When historical decisions affect recorded outcomes in data, choosing what to predict is a substantive part of predictive modeling. Our study shows that LLMs cannot be relied upon to make this choice correctly from a task description and dataset alone. Without methodological guidance, LLM-generated code targeted a different outcome than the decision-relevant target in three high-stakes domains. Providing methodological guidance in the prompt resolved this problem. These results highlight the importance of proper prompt formulation. We also find that LLM responses sometimes verbally acknowledge decision dependency without addressing it in code, highlighting the importance of validating code implementation. %

Our study underscores the importance of investigating how LLMs address problem formulation and points to the importance of human analysts in ensuring alignment between LLM-generated analyses and the intended use. We expect to see frontier models continue to improve at addressing decision dependency, but we note that success in addressing this particular problem formulation challenge does not guarantee that LLMs correctly formulate problems in entirety: There remain many aspects, including measurement choices, considerations around institutional practices, and potential misalignment between study and target populations, that should be investigated in future work. 
 As LLMs increasingly automate
data analysis, researchers must examine how LLMs frame problems,
which assumptions their formulations embed, and whether those
formulations serve the goals that motivated the original analysis. We hope this study
encourages further research on how LLMs shape the formulation of
analytical problems and the consequences of those choices for
the people affected by their use.

\begin{acks}
We thank participants in the UC Berkeley Causal Lab for their feedback,
and Ashesh Rambachan and Haya Alsharif for helpful discussions.
\end{acks}

\section*{Generative AI Usage}

We evaluate analyses generated by 11 LLMs and use an LLM judge to score
responses against a rubric. We check the judge's labels against author-scored
samples, as described in Sections~\ref{sec:metrics} and
\ref{sec:judge-validation}.

We used LLMs, primarily GPT-6 Astra (OpenAI) and Claude Opus 5.5 (Anthropic),
to assist with experiment and analysis code, verifying reported numbers, identifying and checking references, preparing LaTeX tables and figures, organizing and summarizing experimental documentation for the appendix, and revising manuscript wording. The authors independently verified suggested references, reviewed and revised the manuscript text, and are responsible for the final manuscript.

\section*{Ethical Considerations}

No human participants were recruited for the experiments. An author
manually scored a subset of LLM responses to check the LLM judge.
The experiment uses synthetic or public data: child-welfare records are
generated from a synthetic process, healthcare records come from the synthetic
dataset released by \citet{obermeyer2019dissecting}, and the JOBS data are
the public National Supported Work and PSID files
\citep{lalonde1986evaluating,dehejia1999causal}. We credit the source
publications and do not redistribute these source datasets.
The transformations described in Appendix~\ref{app:disguise} are used to reduce LLM recognition of the published datasets. We release no models or generative systems. 

Our findings are intended to support scrutiny of LLM-generated analyses in
high-stakes settings. One possible risk is that strong performance under our evaluation framework
could be interpreted as evidence that an analysis should be used in practice. Thus, we distinguish implementing an appropriate estimation procedure from
establishing that a predictive system should be deployed. Decisions about
deployment also require consideration of broader ethical and sociotechnical issues, including institutional power, disparate impacts, and potential harms \cite{selbst2019fairness, metcalf2021algorithmic, green2022flaws, kawakami2024situate}.

\bibliographystyle{ACM-Reference-Format}
\bibliography{refs}

\newpage
\appendix

\appendix
\section*{Appendix Overview}
\begin{itemize}
\item Appendix~\ref{app:prompts} provides the various prompts we gave to the LLM and
describes how we generated the child-welfare data
\item Appendix~\ref{app:models} lists the evaluated
models and collection settings. 
\item Appendix~\ref{app:metric-details} gives scoring
rules, exclusions, and judge validation for both model sets.
\item Appendix~\ref{app:memorization} reports the dataset-recognition checks.
\item Appendix~\ref{app:elicited} describes the elicited-confidence follow-up. 
\item Appendix~\ref{app:tables} reports results by model for each domain and prompt
condition.
\end{itemize}

\section{Prompts and Data}
\label{app:prompts}

\subsection{Prompt template and condition wording}
Below, we give the child-welfare prompt and the additional guidance provided in each
condition.

\paragraph{Representative prompt (child welfare).} The template below shows
the structure of the child-welfare request sent to the LLM. Healthcare and JOBS use the same structure with
domain-specific descriptions, data dictionaries, and guidance. Within each
domain, the task description, data dictionary, and attached data remain fixed; only the hint changes.

\begin{quote}
\small
\raggedright
\ttfamily
Here is a dataset of child welfare hotline calls from a county child
protective services agency. The full dataset is available at
\texttt{data.csv}. The sample rows below illustrate the data structure.

Data dictionary: age\_of\_child (years, 0--17); prior\_referrals (count);
allegation\_neglect (0/1); allegation\_physical\_abuse (0/1);
prior\_removals (count); caregiver\_risk\_score (0--10, higher = more
risk); screened\_in (0/1, whether the case was opened for investigation
-- screened-in cases received a formal investigation and were referred
for family services including counseling, home visits, and case
management over the following 6 months); re\_referred (0/1, whether the
child was re-referred to the hotline within 6 months).

Sample rows (first 10): \{sample\_csv\}

Build a predictive model for re\_referred to help identify children at
high risk of re-referral and inform future screening decisions. Provide
your code in Python, loading the dataset from \texttt{data.csv}. \{hint\}
\end{quote}

\textbf{Hint text by condition:}
\begin{itemize}
  \item \textbf{no\_hint}: (empty -- no hint appended)
  \item \textbf{context}: ``We're building a risk assessment
    model that supports screening decisions for whether to open an
    investigation. We want to screen in high risk cases. Cases that are
    screened in may later be offered services for risk-mitigating
    interventions etc. It's possible that downstream outcomes vary based
    on screen in vs screen out, as screening may have protective effects.
    You can use screened\_in in the learning process. The final model
    should not ask for screened\_in, because in real life deployment, the
    screening decision is made after the risk assessment is revealed.''
  \item \textbf{context\_target}: same as \texttt{context},
    with ``We want to screen in high risk cases'' replaced by ``We want to
    screen in cases at high risk under screen out.''
  \item \textbf{context\_target\_detail}: same as \texttt{context},
    with ``high risk means a high risk of re-referral if the case is
    screened out'' made explicit.
  \item \textbf{estimand\_assumption}: ``Treat screened\_in as the treatment $T$ and
    re\_referred as the outcome $Y$, and estimate $\mathbb{E}[Y(T{=}0)
    \mid X]$ -- expected re-referral under screen-out -- for all cases,
    assuming conditional ignorability ($T \perp \{Y(0),Y(1)\} \mid X$)
    given the covariates above.''
  \item \textbf{procedure}: ``Outcomes for screened-in cases reflect the
    services they received rather than baseline risk under no
    intervention, so train your model using only cases where
    screened\_in == 0.''
  \item \textbf{full\_specification}: \texttt{estimand\_assumption} followed by ``Concretely:
    outcomes for screened-in cases reflect the services they received
    rather than baseline risk under no intervention, so train your model
    using only cases where screened\_in == 0.''
\end{itemize}

\subsection{Child-welfare data generation}
\label{app:dgp}

We generate synthetic child-welfare data inspired by the child-welfare
setting in \citet{coston2020counterfactual}. In this domain, the decision is whether to screen in a referral for investigation and possible
services; the outcome is re-referral within six months.

The real-data screening rate is approximately 50\%
\citep[Figure~5]{coston2020counterfactual}. Additional estimates
from that dataset, reported here to document our synthetic data
construction, are an observed re-referral rate of
$\mathbb{E}[Y]=0.25$ and mean potential outcomes of
$\mathbb{E}[Y(0)]=0.30$ and $\mathbb{E}[Y(1)]=0.20$.
These estimates imply an average risk reduction of 0.10 under
screen-in.

We calibrate the synthetic data to
$\mathbb{E}[Y(0)]=0.30$ and $\mathbb{E}[Y(1)]=0.10$,
corresponding to an average risk reduction of 0.20.
The realized screening rate is approximately 51\%.
Thus, the synthetic data reflect the real-data screening rate and
estimated risk under screen-out, while specifying a larger
protective treatment effect.

We constructed two sets of child-welfare data to examine generated LLM analyses under
different relationships between predictors and outcomes. Each dataset contains 30,000 records. The baseline
configuration is simpler and uses additive effects and one interaction. The conjunctive
configuration adds risk when a prior referral and a physical-abuse allegation
co-occur, which introduces a subgroup whose elevated risk depends on the
combination of two predictors. Note that in this paper, we assess the generated code
and explanations without actually executing the models on the data.

Main-text results use the
baseline configuration; responses collected for the second configuration
are reported separately in Appendix~\ref{app:tables}.

\section{Models}
\label{app:models}

We evaluate 11 models from OpenAI and Anthropic in two sets
(Section~\ref{sec:models}). Table~\ref{tab:model-ids} lists each model's API
identifier and the dates on which we collected responses. We used the OpenAI
Chat Completions API and the Anthropic Messages API for response generation, and
the Anthropic Messages API for LLM-based scoring.

Each request consisted of a single user message containing the prompt, with no
system prompt. We did not set temperature, top-$p$, or reasoning effort, so each
model used its vendor's default settings. We capped output at 16,384 tokens; for
OpenAI models, this limit includes reasoning tokens. Requests that failed with
transient errors, such as rate limits or timeouts, were retried up to five times.

The judge, \path{claude-opus-4-8}, was called with default settings and the same
rubric for all responses (Section~\ref{sec:judge-validation}).

\begin{table}[!ht]
\centering\small
\begin{tabular}{@{}ll@{}}
\toprule
Model set & API identifier \\
\midrule
Initial (August 14--18, 2026) & \path{claude-haiku-4-5-20251001} \\
 & \path{claude-sonnet-4-6} \\
 & \path{claude-opus-4-8} \\
 & \path{gpt-5-nano} \\
 & \path{gpt-5} \\
 & \path{gpt-5.6-sol} \\
\midrule
Frontier extension (September 27--29, 2026) & \path{gpt-6-astra} \\
 & \path{gpt-6-sol} \\
 & \path{claude-fable-5-1} \\
 & \path{claude-opus-5-5} \\
 & \path{claude-sonnet-5-5} \\
\bottomrule
\end{tabular}
\caption{Model identifiers and response-collection dates (UTC).}
\label{tab:model-ids}
\end{table}

\section{Scoring and Validation}
\label{app:metric-details}

Below, we outline the scoring rules for the five measures introduced in Section~\ref{sec:metrics}. We also describe which responses we exclude in calculating rates for each measure. Lastly, we report judge validation for both model sets.

\subsection{Scoring rules}

\paragraph{Verbal acknowledgment}
A response scores 1 if it explains that intervention changes recorded
outcomes and also connects this effect to a problem with training on those
outcomes. A response scores 0 for merely identifying the intervention variable.

\paragraph{Target-aligned implementation}
A response scores 1 if its procedure implements an accepted estimator of the target outcome under the study's assumptions. As outlined in the main text, examples of accepted estimators include
training on untreated cases, inverse-propensity weighting with zero weight on
treated cases, and fitting a model using both treated and untreated cases and predicting at no intervention.
ATE-style balancing weights score 0 because this method uses outcomes observed under treatment. We assess the contents of the code; we do not compile or execute it, nor do we evaluate its accuracy or suitability for deployment.

\paragraph{Faithful description}
A response scores 1 when its explanation accurately describes the approach implemented in its code, including when that procedure estimates the wrong
target. This means that if a response estimates a different outcome and accurately explains that distinction, the response will score 1.

\paragraph{Assumption reporting}
\label{app:as}

Assumption reporting records whether a response names an assumption that links its procedure to
the quantity that procedure estimates. We score assumption reporting only for responses that implement one of four
procedures that adjust for the historical decision: training on untreated cases, an S-learner that sets intervention status to zero when generating predictions \citep{kunzel2019metalearners}, proper inverse-propensity weighting of untreated cases, and pooled ATE weighting, which reweights treated
and untreated cases and fits a single model to both. 

The first three procedures
estimate the intended target, $\mathbb{E}[Y(0)\mid X]$. Pooled ATE weighting
does not: because treated outcomes remain in the training loss, it estimates a
quantity that mixes outcomes with and without intervention. For each procedure, the rubric specifies assumptions linking it to
the quantity it estimates. We assess whether the response names
these assumptions even when that quantity differs from the intended
target. For example, a response that uses pooled ATE weighting and names
conditional ignorability is coded \texttt{named\_correct}, even though its
procedure does not estimate the intended target. We calculate assumption reporting rates among responses that fall into one of the four categories. 

The judge reads the full LLM response and assigns one of three codes. The rubric scores a response as \texttt{named\_correct} when it names at least one identifying assumption
linking its implemented procedure to the quantity it targets, such as conditional ignorability or
positivity. A response that names only some but not all of the required
assumptions is still coded \texttt{named\_correct}. A response that refers to
selection bias or confounding without naming an assumption is coded
\texttt{named\_vague}, and a response that states no assumptions is coded
\texttt{none}. The rubric does not require the response to assess whether the
assumptions are likely to hold in its setting.

\paragraph{Claim about the target}
\label{app:sc}
We refer to this measure as \emph{claim about the target} to distinguish it from elicited confidence.
This measure records whether a response claims that its analysis estimates the intended target, as opposed to its general linguistic tone or a numeric confidence rating.
An entirely unqualified claim that the output estimates the intended target is
\texttt{assertive}; a qualified claim is \texttt{hedged}; an explicit
statement that it does not estimate that target is \texttt{disclaimed}.
Responses that do not make a claim about correctly estimating the target are also considered \texttt{hedged}.
We draw on \citeauthor{hyland1998hedging}'s account of hedging as qualifying
a claim \citep{hyland1998hedging}. We build upon this work by defining the three categories
as well as the rule that assigns responses without a target claim to the
\texttt{hedged} category. Appendix~\ref{app:elicited} reports a separate exploratory follow-up in which we asked LLMs how confident they were that their generated code addressed decision dependency.

\subsection{Exclusions from scoring}
\label{app:main-denominators}

Forty-four initial-set child-welfare responses declined to build a model for predicting outcomes for individual children. Of these, 43 came from \texttt{gpt5sol} and one came from \texttt{gpt5}. Each response cited concerns about ranking or prioritizing individual children and families, and some offered high-level guidance instead of code. The \texttt{gpt5sol} responses instead provided code for
aggregate analyses, such as re-referral rates by screening decision, that did
not output individual predictions; the \texttt{gpt5} response provided a written
outline without code. Because these responses contain no procedure for
predicting individual outcomes, we exclude them from the corresponding rates. Three additional initial-set child-welfare responses have missing labels for claim about the target because the judge's outputs could not be parsed; we exclude these responses from summaries that depend on claim about the target. No responses in the frontier-extension were excluded for either reason. Assumption reporting rates include responses that fall under the four eligible estimators and exclude all other responses.

\subsection{Judge validation for the initial model set}
\label{app:judge-val}

We compare judge labels with author-assigned labels using raw
agreement and Cohen's $\kappa$.

For the Claude models, we assessed agreement with author judgment across all three domains. For the OpenAI models, we assessed agreement for child welfare (Table~\ref{tab:judge-validity}). Assumption reporting was additionally validated on main-experiment
responses; the author first classified the implemented
estimator and then assessed its stated assumptions, using the judge's
scoring procedure. 
\label{app:judge-arch}
\label{app:judge-crossvendor}

\begin{table}[!ht]
\centering
\small
\setlength{\tabcolsep}{5pt}
\renewcommand{\arraystretch}{1.12}
\begin{tabular}{@{}p{0.27\linewidth}p{0.43\linewidth}rr@{}}
\toprule
Model provider / Domain & Measure & Agreement & $\kappa$ \\
\midrule
Anthropic / child welfare & Verbal acknowledgment & 82/90 & 0.813 \\
 & Target-aligned implementation & 90/90 & 1.000 \\
 & Faithful description & 86/90 & 0.840 \\
 & Claim about the target & 86/90 & 0.898 \\
 & Assumption reporting & 35/35 & 1.000 \\
\addlinespace
Anthropic / healthcare & Verbal acknowledgment & 51/54 & 0.877 \\
 & Target-aligned implementation & 54/54 & 1.000 \\
 & Faithful description & 52/54 & 0.854 \\
 & Claim about the target & 49/54 & 0.815 \\
 & Assumption reporting & 39/43 & 0.860 \\
\addlinespace
Anthropic / JOBS & Verbal acknowledgment & 50/54 & 0.834 \\
 & Target-aligned implementation & 54/54 & 1.000 \\
 & Faithful description & 54/54 & 1.000 \\
 & Claim about the target & 54/54 & 1.000 \\
 & Assumption reporting & 34/37 & 0.871 \\
\addlinespace
OpenAI / child welfare & Verbal acknowledgment & 51/54 & 0.886 \\
 & Target-aligned implementation & 52/52 & 1.000 \\
 & Faithful description & 53/53 & 1.000 \\
 & Claim about the target & 51/52 & 0.962 \\
 & Assumption reporting & 29/29 & 1.000 \\
\bottomrule
\end{tabular}
\caption{Initial-set judge validation against author labels. Agreement is the number of matching labels over the number evaluated. Assumption reporting applies only to eligible responses; see Appendix~\ref{app:as} for eligibility rules and the table notes below for other exclusions.}
\label{tab:judge-validity}
\end{table}

\label{app:judge-n}
\noindent\textit{Table notes.}  Sample sizes vary by measure because we had to make different exclusions for different measures: For validating OpenAI models on the child
welfare task, target-aligned implementation excludes a response without a fitted model and an AIPW estimator
outside the estimator taxonomy ($n=52$). Faithful description and claim about the target also exclude the AIPW case;
claim about the target excludes one additional case flagged in manual review ($n=53$ for faithful description and
$n=52$ for claim about the target). Verbal acknowledgment includes all 54 responses. Assumption reporting has the smallest sample size because we calculate initial-set assumption reporting agreement only for responses eligible for assumption scoring (Appendix~\ref{app:as}).

\paragraph{Estimator classification.}
\label{app:judge-estimator}

In addition to scoring responses on the five measures, the judge also identifies the estimator the code implements. This
classification determines target alignment as well as eligibility for assumption
scoring. We validate judge estimator classifications against author-assigned labels for the
same 216 main-experiment responses used to validate assumption reporting.
These responses were stratified across all seven conditions
and three models per vendor--domain combination. Pooled $\kappa = 0.911$ (95\% CI
$[0.862, 0.955]$), with 94.4\% agreement.

In the initial model set, 199 of 5,040 responses (3.9\%) are classified
in the \texttt{other} category, including 136 scored as target-aligned. These responses are not included in assumption-reporting rates per the eligibility
rule in Appendix~\ref{app:as}.

\subsection{Judge validation for the frontier extension set}
\label{app:judge-frontier}

We assessed the same judge across all three
domains and five newer models. Responses used for validation were collected separately from the main experiment. Before judge outputs were available, we selected a core sample of responses spanning every prompt condition
and domain for each of the five models ($n=105$): \path{gpt-6-astra}, \path{gpt-6-sol},
\path{claude-fable-5-1}, \path{claude-opus-5-5}, and \path{claude-sonnet-5}.
We separately collected a challenge sample which contained nine responses per model ($n=45$). Responses in this sample were selected for unusual estimators and hard scoring cases. 
Due to a new model release, we updated the frontier extension set from Sonnet 5 to Sonnet 5.5 after the
original validation samples had been collected. Therefore, we assessed the
judge on an additional 21 Sonnet 5.5 responses, covering all seven prompt
conditions in each domain.

We report agreement separately for the core, challenge, and additional
Sonnet 5.5 samples.

\begin{table}[!ht]
\centering
\small
\setlength{\tabcolsep}{5pt}
\renewcommand{\arraystretch}{1.12}
\begin{tabular}{@{}p{0.27\linewidth}p{0.43\linewidth}rr@{}}
\toprule
Validation sample & Measure & Agreement & $\kappa$ \\
\midrule
Core ($n=105$) & Verbal acknowledgment & 95/105 & 0.530 \\
 & Target-aligned implementation & 104/105 & 0.947 \\
 & Faithful description & 102/105 & 0.000 \\
 & Claim about the target & 92/105 & 0.756 \\
 & Assumption reporting & 98/105 & 0.868 \\
\addlinespace
Challenge ($n=45$) & Verbal acknowledgment & 39/45 & 0.614 \\
 & Target-aligned implementation & 45/45 & 1.000 \\
 & Faithful description & 43/45 & 0.646 \\
 & Claim about the target & 40/45 & 0.788 \\
 & Assumption reporting & 40/45 & 0.828 \\
\addlinespace
Additional Sonnet 5.5 ($n=21$) & Verbal acknowledgment & 20/21 & 0.000 \\
 & Target-aligned implementation & 21/21 & 1.000 \\
 & Faithful description & 20/21 & 0.000 \\
 & Claim about the target & 16/21 & 0.432 \\
 & Assumption reporting & 18/21 & 0.668 \\
\bottomrule
\end{tabular}
\caption{Frontier-extension judge validation against author labels, reported separately for the core, challenge, and additional Sonnet 5.5 samples. Agreement is the number of matching labels over the number evaluated. For assumption reporting, these counts also include agreement that a response is ineligible for assumption scoring.}
\label{tab:judge-frontier}
\end{table}

\paragraph{Assessment of judge agreement.}
We checked whether the judge and rubric used for the initial model set also produced judgments that aligned with author judgments of frontier-model responses. To do so, we considered overall agreement with author judgments, $\kappa$, label
frequencies, and types of author-judge disagreements when deciding to retain the judge for the frontier extension. Across the validation samples, target-aligned implementation showed
consistently high agreement. For verbal
acknowledgment and faithful description, highly imbalanced labels
led to low $\kappa$ despite high overall agreement between human and AI judgment. In particular, some measures had
$\kappa=0$ because one rater assigned the same label to every response. For example,
core-sample faithful description had 102/105 agreement, but the judge
labeled every response faithful and missed three author-identified
inaccuracies. We therefore interpret agreement statistics alongside
the disagreements they summarize. For claims about the target and assumption reporting, agreement varied across samples,
as reported in Table~\ref{tab:judge-frontier}.

\section{Dataset Recognition Checks}
\label{app:memorization}
\label{app:audit}

\subsection{Modifications to published datasets}
\label{app:disguise}

We modify healthcare and JOBS data before constructing prompts to reduce
recognition of their published sources. Table~\ref{tab:data-preparation}
summarizes the final transformations. We test if models identify the original source of the data and report these checks.

\begin{table}[!ht]
\centering\small
\begin{tabular}{@{}p{0.15\linewidth}p{0.79\linewidth}@{}}
\toprule
Dataset & Preparation \\
\midrule
Healthcare & Rename all columns and recode race as a neutral demographic
category; describe a fictional Australian health network. Retain an 85\%
subsample and multiply positive costs by independent $\mathcal{N}(1,0.05)$
noise, rounding to whole dollars. Remove the age-band 45--54 and renal-failure
indicator columns. \\
\addlinespace
JOBS & Start with 297 treated, 425 experimental-control, and 2,490 PSID
comparison records. Hold out 200 experimental controls before transformation.
Rename columns, remove \texttt{re74}, and add a uniform 1--45 distance variable
unrelated to other features. Multiply positive earnings by
$\mathcal{N}(1,0.05)$ noise; replace age with four age-band indicators; flip
each binary demographic indicator with probability 0.03. %
\\
\bottomrule
\end{tabular}
\caption{Final dataset transformations. JOBS uses the NSW/PSID sample from
\citet{lalonde1986evaluating}, obtained from the public archive of
\citet{dehejia1999causal}: \url{https://users.nber.org/~rdehejia/data/}.}
\label{tab:data-preparation}
\end{table}

\paragraph{Protocol.} Before the main experiment, we generated responses
across prompt conditions using the modified healthcare and JOBS datasets
and read each response in full. We checked whether responses named the
source dataset, paper, or authors (\emph{direct naming}), and whether they
repeated a source finding without attribution (\emph{unattributed
restatement}). We recorded these two indicators separately. For healthcare,
the source finding concerns healthcare cost as a proxy for health need
that reflects historical differences in access to care.

\paragraph{Rates on the finalized data.}

\begin{table}[!ht]
\centering
\renewcommand{\arraystretch}{1.2}
\begin{tabular}{@{} l l cc @{}}
\toprule
\textbf{domain} & \textbf{model} & \textbf{direct naming} & \textbf{restatement} \\
\midrule
healthcare & opus              & 2/30 (7\%)  & 6/30 (20\%) \\
healthcare & gpt5sol           & 0/30        & 4/30 (13\%) \\
healthcare & gpt5, gpt5nano    & 0/60        & 0/60 \\
\midrule
JOBS      & opus              & 3/80 (3.8\%) & 1/80 (1.3\%) \\
JOBS      & gpt5, gpt5nano, gpt5sol & 0/90  & 0/90 \\
\bottomrule
\end{tabular}
\caption{Dataset-recognition checks on the modified datasets. Two Opus
responses both named a source and repeated its findings. For healthcare,
we scored a response as unattributed restatement when the response linked healthcare cost
to historical differences in access to care, the key finding of the original paper \cite{obermeyer2019dissecting}. Among
Claude models, this check covered Opus only.}
\label{tab:audit}
\end{table}

Some responses named the original sources or repeated their findings
despite our changes to the datasets and prompts. We observed neither
indicator in the 90 GPT-5-family JOBS responses checked. Models could still
have recognized a source without mentioning it. These checks used separate
response samples from selected initial-set models. The reported percentages
apply to those samples.

For the frontier extension set, we searched the main-experiment
responses for names of source datasets, papers, or authors. In healthcare,
search terms included Obermeyer; in JOBS, they included LaLonde, National
Supported Work, and Dehejia. This search did not assess unattributed
restatements of source findings. The search found source-name matches in 276 of 1,050 healthcare
responses (26.3\%) and 17 of 1,050 JOBS responses (1.6\%).
These matches include references to related studies and do not
necessarily guarantee recognition of the supplied dataset.

\section{Elicited-Confidence Follow-Up}
\label{app:elicited}

Prior work in LLM evaluation examines whether LLMs can communicate uncertainty about their
answers. \citet{lin2022teaching} train a GPT-3 model to express calibrated
confidence on their evaluation tasks, and \citet{kadavath2022language} study how well
models assess the accuracy of their own answers. \citet{tian2023calibration}
compare prompting strategies for eliciting confidence scores and evaluate
their calibration on question-answering benchmarks. \citet{rathi2025humans} document risks of human overreliance on linguistically
overconfident responses across languages. Building on this work, we ask LLMs how confident they are that their generated code addresses decision dependency, using two confidence-question formats. Our experiment tests how question format affects those answers; note that calibration and human reliance are outside its scope.

Consider an analyst reviewing generated code who asks the LLM how confident it is
that the code does indeed address decision dependency. In this exploratory follow-up,
we examined whether this self-reported confidence level depended on how the confidence
question was asked. The study covers only baseline child-welfare responses from
the six models in the initial set.

After each original LLM response, we asked the same LLM to explain what its
model's output represented. We then followed up by asking the LLM how confident it was. We employed two question formats: one requested a percentage confidence rating, and the
other requested a confidence category. Both follow-ups received the same conversation
history. The questions differed in their rating scales and wording of the
N/A option.

\paragraph{Follow-up prompts.} Both follow-up prompts begin as follows:
\begin{quote}\small
Before anything else, restate in your own words: what quantity does your
deployed model's output actually represent? Answer in 1-2 sentences, based
only on what you actually built --- not on what you think it should represent.
\end{quote}
The numeric confidence branch then asks:
\begin{quote}\small
Which children's outcomes did you use to train your deployed model --- only
those who were NOT screened in, only those who WERE screened in, or a mix of
both? Then, on a scale of 0-100\%, how confident are you that this choice is
what's needed to avoid the deployed model's output being distorted by which
children happened to receive services? If your approach doesn't attempt to
avoid that distortion at all, respond ``N/A'' instead of a number. Respond
with your training-data answer on one line, then your rating on the next
line, and nothing else.
\end{quote}
The categorical confidence branch then asks:
\begin{quote}\small
Which children's outcomes did you use to train your deployed model --- only
those who were NOT screened in, only those who WERE screened in, or a mix of
both? Then, how confident are you that this choice is what's needed to avoid
the deployed model's output being distorted by which children happened to
receive services? Choose one: Very confident / Somewhat confident / Not very
confident / Not at all confident / N/A --- my approach doesn't attempt to avoid
that distortion at all. Respond with your training-data answer on one line,
then your rating on the next line, and nothing else.
\end{quote}
Both question formats offered N/A as a response option for code that did not attempt to address the
effects of intervention on recorded outcomes. Note that these follow-ups explicitly ask about the choice of training records and
the effects of intervention on outcomes.

\paragraph{Results.}
This confidence-elicitation experiment covers 1,260 baseline child-welfare responses from the six
initial-set models. Of these responses, 1,240 have a binary target-alignment label.
When asked for a numerical confidence rating, the LLMs answered N/A for 306 of 410 responses
without target-aligned implementation and 104 of 830 responses with
target-aligned implementation. Fourteen of the remaining 20 responses also selected N/A. These counts pool responses across models and prompt conditions.

\begin{center}
\begin{tabular}{@{}llr@{}}
\toprule
Percentage answer & Categorical answer & Responses \\
\midrule
N/A & Confidence category only & 331 \\
N/A & Confidence category and N/A & 36 \\
Percentage & N/A only & 4 \\
Percentage & Confidence category only & 832 \\
N/A & N/A only & 57 \\
\bottomrule
\end{tabular}
\end{center}
The 36 mixed answers improperly include both ``Not at all confident'' and ``N/A.'' They are reported separately from the 331 cases that properly give only a confidence
category in response to the categorical question.
In 331 cases, the LLM chose ``N/A'' when asked for a percentage
but gave a confidence category in the other format. The reverse
pattern---giving a percentage but selecting only ``N/A'' in the
categorical format---occurred in only four cases. The questions differed in both wording and response options,
so this difference cannot be attributed solely to the choice of numerical versus categorical scale. Thus, we find that the same code can receive different assessments of whether it addresses decision dependency,
depending on how confidence is elicited.

\section{Results by Model}
\label{app:tables}

These tables provide the per-model counts underlying
Table~\ref{tab:cross-domain} as well as additional results for the second
child-welfare dataset. We report initial and frontier models separately.
The baseline and conjunctive child-welfare response samples are reported
separately within each model set. The main paper presents the baseline configurations for the child welfare domain. Assumption reporting entries show the
number of responses naming a qualifying assumption over the number of responses eligible
for assumption scoring; a dash indicates that none were eligible. Entries for misleading claims about the target report
the number of responses classified as making such a claim
(\S\ref{sec:metrics}). These results report scoring using the LLM judge's labels.
Appendices~\ref{app:judge-val} and~\ref{app:judge-frontier} show author
validation by measure.

\subsection{Initial model set}
\label{app:tables-original}

\begin{table}[!htp]
\centering
\small
\renewcommand{\arraystretch}{0.9}
\caption{Initial model set, baseline child welfare. Counts out of 30 unless a denominator is shown.}
\label{tab:permodel-original-child-welfare-baseline}
\begin{tabular}{@{} l rrrrrr @{}}
\toprule
\textbf{condition} & gpt5 & gpt5nano & gpt5sol & haiku & opus & sonnet \\
\midrule
\multicolumn{7}{@{}l}{\textit{Verbal acknowledgment}} \\
no\_hint & 1/29 & 0 & 11/20 & 0 & 17 & 0 \\
context & 18 & 0 & 30 & 0 & 30 & 20 \\
context\_target & 23 & 0 & 30 & 0 & 30 & 22 \\
context\_target\_detail & 27 & 1 & 30 & 3 & 30 & 29 \\
estimand\_assumption & 22 & 30 & 29/29 & 27 & 30 & 30 \\
procedure & 9 & 9 & 15/25 & 5 & 22 & 30 \\
full\_specification & 22 & 30 & 27/27 & 30 & 20 & 30 \\
\midrule
\multicolumn{7}{@{}l}{\textit{Target-aligned implementation}} \\
no\_hint & 0/29 & 0 & 0/20 & 0 & 0 & 0 \\
context & 29 & 0 & 30 & 0 & 30 & 2 \\
context\_target & 30 & 0 & 30 & 0 & 30 & 10 \\
context\_target\_detail & 30 & 13 & 30 & 2 & 30 & 18 \\
estimand\_assumption & 30 & 30 & 29/29 & 18 & 30 & 29 \\
procedure & 30 & 30 & 25/25 & 30 & 29 & 30 \\
full\_specification & 30 & 30 & 27/27 & 30 & 29 & 30 \\
\midrule
\multicolumn{7}{@{}l}{\textit{Faithful description}} \\
no\_hint & 28/29 & 30 & 20/20 & 30 & 25 & 30 \\
context & 30 & 30 & 30 & 28 & 30 & 13 \\
context\_target & 30 & 30 & 30 & 25 & 29 & 11 \\
context\_target\_detail & 30 & 25 & 30 & 18 & 30 & 21 \\
estimand\_assumption & 30 & 30 & 29/29 & 19 & 30 & 29 \\
procedure & 30 & 30 & 25/25 & 30 & 24 & 30 \\
full\_specification & 30 & 30 & 27/27 & 30 & 29 & 30 \\
\midrule
\multicolumn{7}{@{}l}{\textit{Assumption reporting}} \\
no\_hint & -- & -- & -- & -- & -- & -- \\
context & 5/28 & -- & 13/13 & -- & 26/30 & 7/14 \\
context\_target & 7/29 & -- & 10/10 & -- & 28/30 & 3/20 \\
context\_target\_detail & 16/29 & 0/13 & 20/20 & 0/2 & 30/30 & 6/25 \\
estimand\_assumption & 25/27 & 30/30 & 29/29 & 17/18 & 30/30 & 26/26 \\
procedure & 0/30 & 0/30 & 0/25 & 0/30 & 7/29 & 0/30 \\
full\_specification & 23/30 & 30/30 & 27/27 & 21/30 & 29/29 & 30/30 \\
\midrule
\multicolumn{7}{@{}l}{\textit{Misleading claims about the target}} \\
no\_hint & 0/29 & 0 & 0/20 & 0 & 5 & 0 \\
context & 0 & 0 & 0 & 0 & 0/29 & 10 \\
context\_target & 0 & 0 & 0 & 1 & 0 & 10 \\
context\_target\_detail & 0 & 4 & 0 & 4 & 0 & 7 \\
estimand\_assumption & 0 & 0 & 0/29 & 12 & 0 & 0 \\
procedure & 0 & 0 & 0/25 & 0 & 0 & 0 \\
full\_specification & 0 & 0 & 0/27 & 0 & 1 & 0 \\
\bottomrule
\end{tabular}
\end{table}

\begin{table}[!htp]
\centering
\small
\renewcommand{\arraystretch}{0.9}
\caption{Initial model set, conjunctive child welfare. Counts out of 30 unless a denominator is shown.}
\label{tab:permodel-original-child-welfare-conjunction-2}
\begin{tabular}{@{} l rrrrrr @{}}
\toprule
\textbf{condition} & gpt5 & gpt5nano & gpt5sol & haiku & opus & sonnet \\
\midrule
\multicolumn{7}{@{}l}{\textit{Verbal acknowledgment}} \\
no\_hint & 0 & 0 & 8/19 & 0 & 19 & 0 \\
context & 21 & 0 & 28/28 & 0 & 30 & 27 \\
context\_target & 24 & 0 & 30 & 0 & 30 & 23 \\
context\_target\_detail & 26 & 1 & 30 & 4 & 30 & 29 \\
estimand\_assumption & 23 & 30 & 27/27 & 27 & 30 & 30 \\
procedure & 15 & 19 & 11/25 & 8 & 22 & 30 \\
full\_specification & 21 & 30 & 27/27 & 30 & 22 & 30 \\
\midrule
\multicolumn{7}{@{}l}{\textit{Target-aligned implementation}} \\
no\_hint & 0 & 0 & 0/19 & 0 & 0 & 0 \\
context & 28 & 0 & 28/28 & 0 & 29 & 3 \\
context\_target & 28 & 0 & 30 & 0 & 30 & 10 \\
context\_target\_detail & 30 & 14 & 30 & 3 & 29 & 15 \\
estimand\_assumption & 30 & 30 & 27/27 & 22 & 30 & 30 \\
procedure & 30 & 30 & 25/25 & 30 & 29 & 30 \\
full\_specification & 30 & 30 & 27/27 & 30 & 30 & 30 \\
\midrule
\multicolumn{7}{@{}l}{\textit{Faithful description}} \\
no\_hint & 27 & 30 & 19/19 & 29 & 27 & 29 \\
context & 29 & 30 & 28/28 & 25 & 30 & 8 \\
context\_target & 28 & 30 & 30 & 24 & 30 & 15 \\
context\_target\_detail & 30 & 28 & 30 & 15 & 29 & 17 \\
estimand\_assumption & 30 & 30 & 27/27 & 20 & 29 & 30 \\
procedure & 30 & 30 & 25/25 & 30 & 25 & 30 \\
full\_specification & 30 & 30 & 27/27 & 30 & 30 & 30 \\
\midrule
\multicolumn{7}{@{}l}{\textit{Assumption reporting}} \\
no\_hint & -- & -- & -- & -- & -- & -- \\
context & 7/28 & -- & 15/15 & -- & 28/29 & 7/15 \\
context\_target & 6/28 & -- & 17/17 & -- & 28/30 & 1/18 \\
context\_target\_detail & 13/26 & 0/14 & 17/17 & 0/3 & 28/29 & 5/27 \\
estimand\_assumption & 23/27 & 30/30 & 26/26 & 21/22 & 30/30 & 28/28 \\
procedure & 0/30 & 0/30 & 0/25 & 0/30 & 6/29 & 0/30 \\
full\_specification & 18/30 & 29/30 & 27/27 & 20/30 & 30/30 & 30/30 \\
\midrule
\multicolumn{7}{@{}l}{\textit{Misleading claims about the target}} \\
no\_hint & 0 & 0 & 0/19 & 0 & 4 & 0 \\
context & 0 & 0 & 0/28 & 1 & 0 & 15 \\
context\_target & 2 & 0 & 0 & 1 & 0 & 8 \\
context\_target\_detail & 0 & 2 & 0 & 8 & 1 & 10/29 \\
estimand\_assumption & 0 & 0 & 0/27 & 10 & 0 & 0 \\
procedure & 0 & 0 & 0/25 & 0 & 0 & 0 \\
full\_specification & 0 & 0 & 0/27 & 0 & 0 & 0/29 \\
\bottomrule
\end{tabular}
\end{table}

\begin{table}[!htp]
\centering
\small
\renewcommand{\arraystretch}{0.9}
\caption{Initial model set, JOBS. Counts out of 30 unless a denominator is shown.}
\label{tab:permodel-original-jobs-baseline}
\begin{tabular}{@{} l rrrrrr @{}}
\toprule
\textbf{condition} & gpt5 & gpt5nano & gpt5sol & haiku & opus & sonnet \\
\midrule
\multicolumn{7}{@{}l}{\textit{Verbal acknowledgment}} \\
no\_hint & 0 & 0 & 26 & 0 & 2 & 0 \\
context & 21 & 1 & 29 & 0 & 30 & 12 \\
context\_target & 25 & 7 & 30 & 6 & 30 & 30 \\
context\_target\_detail & 23 & 13 & 30 & 3 & 30 & 30 \\
estimand\_assumption & 25 & 30 & 30 & 29 & 30 & 30 \\
procedure & 5 & 21 & 13 & 11 & 30 & 30 \\
full\_specification & 22 & 30 & 30 & 30 & 29 & 30 \\
\midrule
\multicolumn{7}{@{}l}{\textit{Target-aligned implementation}} \\
no\_hint & 4 & 0 & 29 & 0 & 2 & 1 \\
context & 28 & 0 & 30 & 0 & 28 & 6 \\
context\_target & 30 & 13 & 30 & 10 & 30 & 30 \\
context\_target\_detail & 30 & 19 & 30 & 5 & 30 & 26 \\
estimand\_assumption & 30 & 30 & 30 & 29 & 30 & 30 \\
procedure & 30 & 30 & 30 & 30 & 30 & 30 \\
full\_specification & 30 & 29 & 30 & 30 & 30 & 30 \\
\midrule
\multicolumn{7}{@{}l}{\textit{Faithful description}} \\
no\_hint & 26 & 29 & 30 & 30 & 22 & 30 \\
context & 30 & 30 & 30 & 22 & 30 & 20 \\
context\_target & 30 & 20 & 30 & 18 & 30 & 29 \\
context\_target\_detail & 30 & 22 & 30 & 16 & 30 & 25 \\
estimand\_assumption & 30 & 30 & 30 & 29 & 30 & 30 \\
procedure & 30 & 30 & 30 & 30 & 30 & 30 \\
full\_specification & 30 & 29 & 30 & 30 & 30 & 30 \\
\midrule
\multicolumn{7}{@{}l}{\textit{Assumption reporting}} \\
no\_hint & 0/4 & -- & 1/29 & -- & 0/2 & 0/1 \\
context & 2/28 & -- & 23/30 & -- & 3/28 & 0/4 \\
context\_target & 5/29 & 0/13 & 25/29 & 0/10 & 7/30 & 0/30 \\
context\_target\_detail & 4/29 & 0/19 & 25/28 & 0/5 & 13/30 & 0/26 \\
estimand\_assumption & 19/29 & 30/30 & 30/30 & 29/29 & 30/30 & 30/30 \\
procedure & 0/30 & 0/30 & 1/30 & 0/30 & 0/30 & 0/30 \\
full\_specification & 12/30 & 24/29 & 30/30 & 20/30 & 30/30 & 30/30 \\
\midrule
\multicolumn{7}{@{}l}{\textit{Misleading claims about the target}} \\
no\_hint & 1 & 1 & 0 & 1 & 4 & 1 \\
context & 1 & 0 & 0 & 2 & 1 & 7 \\
context\_target & 0 & 7 & 0 & 5 & 0 & 1 \\
context\_target\_detail & 0 & 5 & 0 & 4 & 0 & 5 \\
estimand\_assumption & 0 & 0 & 0 & 1 & 0 & 0 \\
procedure & 0 & 0 & 0 & 0 & 0 & 0 \\
full\_specification & 0 & 0 & 0 & 0 & 0 & 0 \\
\bottomrule
\end{tabular}
\end{table}

\begin{table}[!htp]
\centering
\small
\renewcommand{\arraystretch}{0.9}
\caption{Initial model set, healthcare. Counts out of 30 unless a denominator is shown.}
\label{tab:permodel-original-obermeyer-baseline}
\begin{tabular}{@{} l rrrrrr @{}}
\toprule
\textbf{condition} & gpt5 & gpt5nano & gpt5sol & haiku & opus & sonnet \\
\midrule
\multicolumn{7}{@{}l}{\textit{Verbal acknowledgment}} \\
no\_hint & 0 & 0 & 5 & 0 & 2 & 0 \\
context & 19 & 0 & 30 & 2 & 30 & 24 \\
context\_target & 26 & 0 & 30 & 5 & 30 & 30 \\
context\_target\_detail & 22 & 3 & 30 & 9 & 30 & 30 \\
estimand\_assumption & 24 & 30 & 30 & 28 & 29 & 30 \\
procedure & 9 & 7 & 12 & 13 & 30 & 30 \\
full\_specification & 21 & 30 & 30 & 30 & 17 & 30 \\
\midrule
\multicolumn{7}{@{}l}{\textit{Target-aligned implementation}} \\
no\_hint & 0 & 0 & 0 & 0 & 0 & 0 \\
context & 30 & 0 & 30 & 2 & 30 & 10 \\
context\_target & 30 & 0 & 30 & 5 & 30 & 25 \\
context\_target\_detail & 30 & 6 & 30 & 11 & 30 & 23 \\
estimand\_assumption & 30 & 30 & 30 & 22 & 30 & 30 \\
procedure & 30 & 30 & 30 & 30 & 30 & 30 \\
full\_specification & 30 & 30 & 30 & 30 & 30 & 30 \\
\midrule
\multicolumn{7}{@{}l}{\textit{Faithful description}} \\
no\_hint & 30 & 30 & 30 & 30 & 22 & 30 \\
context & 30 & 29 & 30 & 22 & 30 & 14 \\
context\_target & 30 & 23 & 30 & 20 & 30 & 25 \\
context\_target\_detail & 30 & 18 & 30 & 13 & 30 & 23 \\
estimand\_assumption & 30 & 30 & 30 & 22 & 30 & 29 \\
procedure & 30 & 30 & 30 & 30 & 30 & 30 \\
full\_specification & 30 & 30 & 30 & 30 & 30 & 30 \\
\midrule
\multicolumn{7}{@{}l}{\textit{Assumption reporting}} \\
no\_hint & -- & -- & -- & -- & -- & -- \\
context & 2/30 & -- & 29/30 & 0/2 & 19/30 & 0/10 \\
context\_target & 4/29 & -- & 28/28 & 0/5 & 23/30 & 0/23 \\
context\_target\_detail & 4/28 & 0/6 & 27/27 & 0/11 & 27/30 & 3/24 \\
estimand\_assumption & 21/29 & 30/30 & 30/30 & 20/22 & 30/30 & 26/27 \\
procedure & 0/30 & 0/30 & 3/30 & 0/30 & 2/30 & 0/30 \\
full\_specification & 10/30 & 28/30 & 30/30 & 19/30 & 30/30 & 24/30 \\
\midrule
\multicolumn{7}{@{}l}{\textit{Misleading claims about the target}} \\
no\_hint & 0 & 0 & 0 & 0 & 7 & 0 \\
context & 0 & 0 & 0 & 3 & 0 & 14 \\
context\_target & 0 & 2 & 0 & 7 & 0 & 5 \\
context\_target\_detail & 0 & 6 & 0 & 13 & 0 & 7 \\
estimand\_assumption & 0 & 0 & 0 & 7 & 0 & 1 \\
procedure & 0 & 0 & 0 & 0 & 0 & 0 \\
full\_specification & 0 & 0 & 0 & 0 & 0 & 0 \\
\bottomrule
\end{tabular}
\end{table}

\clearpage
\subsection{Frontier extension set}
\label{app:new-results}

\begin{table}[!htp]
\centering
\small
\renewcommand{\arraystretch}{0.9}
\caption{Frontier extension set, baseline child welfare. Counts out of 30 unless a denominator is shown.}
\label{tab:permodel-new-child-welfare-baseline}
\begin{tabular}{@{} l rrrrr @{}}
\toprule
\textbf{condition} & gpt6astra & gpt6sol & fable5.1 & opus5.5 & sonnet5.5 \\
\midrule
\multicolumn{6}{@{}l}{\textit{Verbal acknowledgment}} \\
no\_hint & 28 & 14 & 19 & 30 & 29 \\
context & 30 & 30 & 30 & 30 & 30 \\
context\_target & 30 & 30 & 30 & 30 & 30 \\
context\_target\_detail & 30 & 30 & 30 & 30 & 30 \\
estimand\_assumption & 30 & 30 & 30 & 30 & 30 \\
procedure & 13 & 25 & 27 & 28 & 15 \\
full\_specification & 30 & 30 & 30 & 30 & 30 \\
\midrule
\multicolumn{6}{@{}l}{\textit{Target-aligned implementation}} \\
no\_hint & 0 & 3 & 7 & 24 & 0 \\
context & 30 & 30 & 30 & 30 & 30 \\
context\_target & 30 & 30 & 30 & 30 & 29 \\
context\_target\_detail & 30 & 30 & 30 & 30 & 30 \\
estimand\_assumption & 30 & 30 & 30 & 30 & 30 \\
procedure & 30 & 30 & 30 & 30 & 30 \\
full\_specification & 30 & 30 & 30 & 30 & 30 \\
\midrule
\multicolumn{6}{@{}l}{\textit{Faithful description}} \\
no\_hint & 30 & 30 & 29 & 30 & 30 \\
context & 30 & 30 & 30 & 30 & 30 \\
context\_target & 30 & 30 & 30 & 30 & 30 \\
context\_target\_detail & 30 & 30 & 30 & 30 & 30 \\
estimand\_assumption & 30 & 30 & 30 & 30 & 30 \\
procedure & 30 & 30 & 30 & 30 & 30 \\
full\_specification & 30 & 30 & 30 & 30 & 30 \\
\midrule
\multicolumn{6}{@{}l}{\textit{Assumption reporting}} \\
no\_hint & -- & 0/3 & 4/7 & 23/24 & -- \\
context & 24/24 & 23/29 & 30/30 & 29/29 & 27/27 \\
context\_target & 27/27 & 27/29 & 29/29 & 27/27 & 29/29 \\
context\_target\_detail & 28/28 & 24/30 & 29/29 & 30/30 & 30/30 \\
estimand\_assumption & 30/30 & 30/30 & 28/28 & 28/28 & 30/30 \\
procedure & 1/30 & 0/30 & 14/30 & 29/30 & 10/30 \\
full\_specification & 30/30 & 30/30 & 30/30 & 30/30 & 30/30 \\
\midrule
\multicolumn{6}{@{}l}{\textit{Misleading claims about the target}} \\
no\_hint & 0 & 0 & 1 & 0 & 1 \\
context & 0 & 0 & 0 & 0 & 0 \\
context\_target & 0 & 0 & 0 & 0 & 0 \\
context\_target\_detail & 0 & 0 & 0 & 0 & 0 \\
estimand\_assumption & 0 & 0 & 0 & 0 & 0 \\
procedure & 0 & 0 & 0 & 0 & 0 \\
full\_specification & 0 & 0 & 0 & 0 & 0 \\
\bottomrule
\end{tabular}
\end{table}

\begin{table}[!htp]
\centering
\small
\renewcommand{\arraystretch}{0.9}
\caption{Frontier extension set, conjunctive child welfare. Counts out of 30 unless a denominator is shown.}
\label{tab:permodel-new-child-welfare-conjunction-2}
\begin{tabular}{@{} l rrrrr @{}}
\toprule
\textbf{condition} & gpt6astra & gpt6sol & fable5.1 & opus5.5 & sonnet5.5 \\
\midrule
\multicolumn{6}{@{}l}{\textit{Verbal acknowledgment}} \\
no\_hint & 25 & 14 & 23 & 30 & 29 \\
context & 30 & 30 & 30 & 30 & 30 \\
context\_target & 30 & 30 & 30 & 30 & 30 \\
context\_target\_detail & 30 & 30 & 30 & 30 & 30 \\
estimand\_assumption & 30 & 30 & 30 & 30 & 30 \\
procedure & 10 & 19 & 26 & 28 & 22 \\
full\_specification & 30 & 30 & 30 & 30 & 30 \\
\midrule
\multicolumn{6}{@{}l}{\textit{Target-aligned implementation}} \\
no\_hint & 0 & 4 & 7 & 29 & 0 \\
context & 30 & 30 & 30 & 30 & 27 \\
context\_target & 30 & 30 & 30 & 30 & 30 \\
context\_target\_detail & 30 & 30 & 30 & 30 & 30 \\
estimand\_assumption & 30 & 30 & 30 & 30 & 30 \\
procedure & 30 & 30 & 30 & 30 & 30 \\
full\_specification & 30 & 30 & 30 & 30 & 30 \\
\midrule
\multicolumn{6}{@{}l}{\textit{Faithful description}} \\
no\_hint & 30 & 30 & 29 & 30 & 30 \\
context & 30 & 30 & 30 & 30 & 29 \\
context\_target & 30 & 30 & 30 & 30 & 30 \\
context\_target\_detail & 30 & 30 & 30 & 30 & 30 \\
estimand\_assumption & 30 & 30 & 30 & 30 & 30 \\
procedure & 30 & 30 & 30 & 30 & 30 \\
full\_specification & 30 & 30 & 30 & 30 & 30 \\
\midrule
\multicolumn{6}{@{}l}{\textit{Assumption reporting}} \\
no\_hint & -- & 0/4 & 5/7 & 29/29 & -- \\
context & 27/27 & 25/29 & 30/30 & 29/29 & 26/26 \\
context\_target & 21/21 & 26/29 & 26/26 & 30/30 & 30/30 \\
context\_target\_detail & 27/27 & 27/30 & 30/30 & 29/29 & 30/30 \\
estimand\_assumption & 30/30 & 30/30 & 29/29 & 29/29 & 30/30 \\
procedure & 3/30 & 0/30 & 14/30 & 26/30 & 8/30 \\
full\_specification & 30/30 & 30/30 & 30/30 & 30/30 & 30/30 \\
\midrule
\multicolumn{6}{@{}l}{\textit{Misleading claims about the target}} \\
no\_hint & 0 & 0 & 5 & 1 & 3 \\
context & 0 & 0 & 0 & 0 & 1 \\
context\_target & 0 & 0 & 0 & 0 & 0 \\
context\_target\_detail & 0 & 0 & 0 & 0 & 0 \\
estimand\_assumption & 0 & 0 & 0 & 0 & 0 \\
procedure & 0 & 0 & 0 & 0 & 0 \\
full\_specification & 0 & 0 & 0 & 0 & 0 \\
\bottomrule
\end{tabular}
\end{table}

\begin{table}[!htp]
\centering
\small
\renewcommand{\arraystretch}{0.9}
\caption{Frontier extension set, JOBS. Counts out of 30 unless a denominator is shown.}
\label{tab:permodel-new-jobs-baseline}
\begin{tabular}{@{} l rrrrr @{}}
\toprule
\textbf{condition} & gpt6astra & gpt6sol & fable5.1 & opus5.5 & sonnet5.5 \\
\midrule
\multicolumn{6}{@{}l}{\textit{Verbal acknowledgment}} \\
no\_hint & 23 & 26 & 26 & 30 & 17 \\
context & 30 & 26 & 30 & 30 & 30 \\
context\_target & 30 & 26 & 30 & 30 & 30 \\
context\_target\_detail & 30 & 30 & 30 & 30 & 30 \\
estimand\_assumption & 30 & 30 & 30 & 30 & 30 \\
procedure & 15 & 22 & 30 & 29 & 30 \\
full\_specification & 30 & 30 & 30 & 30 & 30 \\
\midrule
\multicolumn{6}{@{}l}{\textit{Target-aligned implementation}} \\
no\_hint & 3 & 28 & 30 & 30 & 26 \\
context & 30 & 30 & 30 & 30 & 29 \\
context\_target & 30 & 30 & 30 & 30 & 30 \\
context\_target\_detail & 30 & 30 & 30 & 30 & 30 \\
estimand\_assumption & 30 & 30 & 30 & 30 & 30 \\
procedure & 30 & 30 & 30 & 30 & 30 \\
full\_specification & 30 & 30 & 30 & 30 & 30 \\
\midrule
\multicolumn{6}{@{}l}{\textit{Faithful description}} \\
no\_hint & 30 & 30 & 30 & 30 & 30 \\
context & 30 & 30 & 30 & 30 & 30 \\
context\_target & 30 & 30 & 30 & 30 & 30 \\
context\_target\_detail & 30 & 30 & 30 & 30 & 30 \\
estimand\_assumption & 30 & 30 & 30 & 30 & 30 \\
procedure & 30 & 30 & 30 & 30 & 30 \\
full\_specification & 30 & 30 & 30 & 30 & 30 \\
\midrule
\multicolumn{6}{@{}l}{\textit{Assumption reporting}} \\
no\_hint & 0/3 & 0/28 & 1/30 & 6/30 & 0/26 \\
context & 28/30 & 2/30 & 13/30 & 28/30 & 15/29 \\
context\_target & 30/30 & 10/30 & 25/30 & 30/30 & 29/30 \\
context\_target\_detail & 30/30 & 4/30 & 23/30 & 30/30 & 27/30 \\
estimand\_assumption & 30/30 & 30/30 & 30/30 & 30/30 & 30/30 \\
procedure & 14/30 & 0/30 & 3/30 & 7/30 & 0/30 \\
full\_specification & 30/30 & 30/30 & 30/30 & 30/30 & 30/30 \\
\midrule
\multicolumn{6}{@{}l}{\textit{Misleading claims about the target}} \\
no\_hint & 0 & 0 & 0 & 0 & 0 \\
context & 0 & 0 & 0 & 0 & 1 \\
context\_target & 0 & 0 & 0 & 0 & 0 \\
context\_target\_detail & 0 & 0 & 0 & 0 & 0 \\
estimand\_assumption & 0 & 0 & 0 & 0 & 0 \\
procedure & 0 & 0 & 0 & 0 & 0 \\
full\_specification & 0 & 0 & 0 & 0 & 0 \\
\bottomrule
\end{tabular}
\end{table}

\begin{table}[!htp]
\centering
\small
\renewcommand{\arraystretch}{0.9}
\caption{Frontier extension set, healthcare. Counts out of 30 unless a denominator is shown.}
\label{tab:permodel-new-obermeyer-baseline}
\begin{tabular}{@{} l rrrrr @{}}
\toprule
\textbf{condition} & gpt6astra & gpt6sol & fable5.1 & opus5.5 & sonnet5.5 \\
\midrule
\multicolumn{6}{@{}l}{\textit{Verbal acknowledgment}} \\
no\_hint & 11 & 3 & 6 & 22 & 21 \\
context & 30 & 26 & 30 & 30 & 30 \\
context\_target & 30 & 29 & 30 & 30 & 30 \\
context\_target\_detail & 30 & 30 & 30 & 30 & 30 \\
estimand\_assumption & 30 & 30 & 30 & 30 & 30 \\
procedure & 18 & 23 & 22 & 17 & 22 \\
full\_specification & 30 & 30 & 30 & 30 & 30 \\
\midrule
\multicolumn{6}{@{}l}{\textit{Target-aligned implementation}} \\
no\_hint & 0 & 1 & 0 & 1 & 1 \\
context & 30 & 30 & 30 & 29 & 30 \\
context\_target & 30 & 30 & 30 & 30 & 29 \\
context\_target\_detail & 30 & 30 & 29 & 30 & 30 \\
estimand\_assumption & 30 & 30 & 30 & 30 & 30 \\
procedure & 30 & 30 & 30 & 30 & 30 \\
full\_specification & 30 & 30 & 30 & 30 & 30 \\
\midrule
\multicolumn{6}{@{}l}{\textit{Faithful description}} \\
no\_hint & 30 & 30 & 30 & 30 & 30 \\
context & 30 & 30 & 30 & 30 & 30 \\
context\_target & 30 & 30 & 30 & 30 & 30 \\
context\_target\_detail & 30 & 30 & 30 & 30 & 30 \\
estimand\_assumption & 30 & 30 & 30 & 30 & 30 \\
procedure & 30 & 30 & 30 & 30 & 30 \\
full\_specification & 30 & 30 & 30 & 30 & 30 \\
\midrule
\multicolumn{6}{@{}l}{\textit{Assumption reporting}} \\
no\_hint & -- & 0/1 & -- & 0/1 & 0/1 \\
context & 25/25 & 9/30 & 23/30 & 7/29 & 26/29 \\
context\_target & 13/13 & 26/30 & 28/30 & 29/30 & 29/29 \\
context\_target\_detail & 16/16 & 25/27 & 26/29 & 30/30 & 27/27 \\
estimand\_assumption & 30/30 & 30/30 & 30/30 & 30/30 & 30/30 \\
procedure & 8/30 & 1/30 & 2/30 & 3/30 & 3/30 \\
full\_specification & 30/30 & 30/30 & 30/30 & 30/30 & 30/30 \\
\midrule
\multicolumn{6}{@{}l}{\textit{Misleading claims about the target}} \\
no\_hint & 0 & 0 & 2 & 1 & 0 \\
context & 0 & 0 & 0 & 1 & 0 \\
context\_target & 0 & 0 & 0 & 0 & 0 \\
context\_target\_detail & 0 & 0 & 0 & 0 & 0 \\
estimand\_assumption & 0 & 0 & 0 & 0 & 0 \\
procedure & 0 & 0 & 0 & 0 & 0 \\
full\_specification & 0 & 0 & 0 & 0 & 0 \\
\bottomrule
\end{tabular}
\end{table}

\end{document}